\documentstyle[epsfig,onecolumn]{mn}

\newif\ifAMStwofonts

\newcommand{\sm}{\, {\rm M}_{\odot}}
\newcommand{\kms}{${\rm km \, s^{-1}}$}
\newcommand{\D}{\displaystyle}

\ifoldfss
  \ifCUPmtlplainloaded \else
    \NewTextAlphabet{textbfit} {cmbxti10} {}
    \NewTextAlphabet{textbfss} {cmssbx10} {}
    \NewMathAlphabet{mathbfit} {cmbxti10} {} 
    \NewMathAlphabet{mathbfss} {cmssbx10} {} 
  \fi
  \ifAMStwofonts
    \ifCUPmtlplainloaded \else
      \NewSymbolFont{upmath} {eurm10}
      \NewSymbolFont{AMSa} {msam10}
      \NewMathSymbol{\upi}     {0}{upmath}{19}
      \NewMathSymbol{\umu}     {0}{upmath}{16}
      \NewMathSymbol{\upartial}{0}{upmath}{40}
      \NewMathSymbol{\leqslant}{3}{AMSa}{36}
      \NewMathSymbol{\geqslant}{3}{AMSa}{3E}

       \let\le=\leqslant
       
    \fi
  \fi
\fi 

\ifnfssone
  \newmathalphabet{\mathit}
  \addtoversion{normal}{\mathit}{cmr}{m}{it}
  \addtoversion{bold}{\mathit}{cmr}{bx}{it}
  \newmathalphabet{\mathbfit} 
  \addtoversion{normal}{\mathbfit}{cmr}{bx}{it}
  \addtoversion{bold}{\mathbfit}{cmr}{bx}{it}
  \newmathalphabet{\mathbfss} 
  \addtoversion{normal}{\mathbfss}{cmss}{bx}{n}
  \addtoversion{bold}{\mathbfss}{cmss}{bx}{n}
  \ifAMStwofonts
    \ifCUPmtlplainloaded \else
      \UseAMStwoboldmath
      \makeatletter
      \new@mathgroup\upmath@group
      \define@mathgroup\mv@normal\upmath@group{eur}{m}{n}
      \define@mathgroup\mv@bold\upmath@group{eur}{b}{n}
      \edef\UPM{\hexnumber\upmath@group}
      \new@mathgroup\amsa@group
      \define@mathgroup\mv@normal\amsa@group{msa}{m}{n}
      \define@mathgroup\mv@bold\amsa@group{msa}{m}{n}
      \edef\AMSa{\hexnumber\amsa@group}
      \makeatother
      \mathchardef\upi="0\UPM19
      \mathchardef\umu="0\UPM16
      \mathchardef\upartial="0\UPM40
      \mathchardef\leqslant="3\AMSa36
      \mathchardef\geqslant="3\AMSa3E

       \let\le=\leqslant

    \fi
  \fi
\fi 

\ifnfsstwo
  \DeclareMathAlphabet{\mathbfit}{OT1}{cmr}{bx}{it}
  \SetMathAlphabet\mathbfit{bold}{OT1}{cmr}{bx}{it}
  \DeclareMathAlphabet{\mathbfss}{OT1}{cmss}{bx}{n}
  \SetMathAlphabet\mathbfss{bold}{OT1}{cmss}{bx}{n}
  \ifAMStwofonts
    \ifCUPmtlplainloaded \else
      \DeclareSymbolFont{UPM}{U}{eur}{m}{n}
      \SetSymbolFont{UPM}{bold}{U}{eur}{b}{n}
      \DeclareSymbolFont{AMSa}{U}{msa}{m}{n}
      \DeclareMathSymbol{\upi}{0}{UPM}{"19}
      \DeclareMathSymbol{\umu}{0}{UPM}{"16}
      \DeclareMathSymbol{\upartial}{0}{UPM}{"40}
      \DeclareMathSymbol{\leqslant}{3}{AMSa}{"36}
      \DeclareMathSymbol{\geqslant}{3}{AMSa}{"3E}

       \let\le=\leqslant

    \fi
  \fi
\fi 

\ifCUPmtlplainloaded \else
  \ifAMStwofonts \else 
    \def\upi{\pi}
    \def\umu{\mu}
    \def\upartial{\partial}
  \fi
\fi

\title{Building up the Stellar Halo of the Galaxy}
\author[A. Helmi and S.D.M. White]
       {Amina Helmi$^{1}$ and Simon D.M. White$^{2}$\\
       $^{1}$ Sterrewacht Leiden, Postbus 9513, 2300 RA Leiden, 
The Netherlands\\
	$^{2}$ Max-Planck-Institut f\"ur Astrophysik, Karl-Schwarzschild-Str. 
1, 85740 Garching bei M\"unchen, Germany}
\date{Accepted ...
      Received ...;
      in original form ...}

\pagerange{\pageref{firstpage}--\pageref{lastpage}}
\pubyear{}

\begin{document}

\maketitle

\label{firstpage}

\begin{abstract}
We study numerical simulations of satellite galaxy disruption
in a potential resembling that of the Milky Way. Our goal is
to assess whether a merger origin for the stellar halo would 
leave observable fossil structure in the phase-space distribution 
of nearby stars. We show how mixing of disrupted satellites can
be quantified using a coarse-grained entropy. Although after 10
Gyr few obvious asymmetries remain in the distribution of 
particles in configuration space, strong correlations are still present
in velocity space. We give a simple analytic description of these 
effects, based on a linearized treatment in action-angle variables,
which shows how the kinematic and density structure of the debris
stream changes with time. By applying this description 
we find that a single dwarf elliptical-like satellite of 
current luminosity $10^8~{\rm L}_\odot$
disrupted 10 Gyr ago from an orbit circulating in the  
inner halo (mean apocentre $\sim 12$ kpc) would contribute about 
$\sim 30$ kinematically cold streams with internal velocity dispersions
below  5 \kms~ to the local stellar halo. 
If the whole stellar halo were built by such
disrupted satellites, it should consist locally of 
$300 - 500$ such streams. Clear detection of all
these structures would require 
a sample of a few thousand stars with 3-D velocities accurate to
better than 5 \kms. Even with velocity errors several times worse than
this, the expected clumpiness should be quite evident. 
We apply our 
formalism to a group of stars detected near the North Galactic Pole, and 
derive an order of magnitude estimate
for the initial properties of the progenitor system.
\end{abstract}

\begin{keywords}
Galaxy: halo, formation, dynamics -- 
galaxies: formation, halos, interactions
\end{keywords}

\section{Introduction}

There have been two different traditional views on the 
formation history of the Milky Way. 
The first model was introduced by Eggen, Lynden-Bell 
\& Sandage (1962) to explain the kinematics 
of metal poor halo field stars in the solar neighbourhood. 
According to their view the Galaxy formed in a monolithic way, 
by the free fall collapse of a relatively 
uniform, star-forming cloud. After the system became rotationally 
supported, further star formation took  place in a metal-enriched
disk,  thereby producing a correlation between kinematics and 
metallicity: the well-known disk-halo transition. 
In later studies Searle \& Zinn (1978) noted the lack of
an abundance gradient and a substantial spread 
in ages in the outer halo globular cluster system.
This led them to propose an alternative picture
in which our Galaxy's stellar halo formed in a more chaotic way through 
merging of several protogalactic clouds. 
(See Freeman 1987 for a complete review). 

This second model resembles more closely the view of the current 
cosmological theories of structure formation in the Universe. 
These theories postulate that structure grows through the 
amplification by the gravitational forces of initially 
small density fluctuations (Peebles 1970; White 1976; Peebles 1980, 1993). 
In all currently popular versions small objects are the first 
to collapse; they then merge forming progressively larger systems 
giving rise to the complex structure of galaxies and galaxy clusters 
we observe today. This hierarchical scenario is currently 
the only well-studied model which places 
galaxy formation in its proper cosmological context
 (see White 1996 for a comprehensive review). 
Numerical simulations of large-scale structure formation show a
remarkable similarity to observational surveys 
(e.g. Jenkins et al. 1997, and references therein; and 
Efstathiou 1996 for a review). For galaxy formation, 
the combination of numerical and semi-analytic modelling
has proved to be very powerful, despite the necessarily schematic
representation of a number of processes affecting the formation of a
galaxy (Katz 1992; Kauffmann, White \& Guiderdoni 1993;  
Cole et al. 1994; Navarro \& White 1994; Steinmetz \& Muller 1995; 
Kauffmann 1996; Mo, Mao \& White 1998; 
Somerville \& Primack 1999; Steinmetz \& Navarro 1999). 
This general framework, where structure forms bottom-up, 
provides the background for our work.

We are motivated, however, not only by this theoretical 
modelling, but also by the 
 increasing number of observations which suggest
substructure in the
halo of the Galaxy (Eggen 1962; Rodgers, Harding \& Sadler 1981; 
Rodgers \& Paltoglou 1984; Ratnatunga \& Freeman 1985; 
Sommer-Larsen \& Christensen 1987; 
Doinidis \& Beers 1989; Arnold \& Gilmore 1992; Preston, 
Beers \& Shectman 1994; Majewski, Munn \& Hawley 1994; 
Majewski, Munn \& Hawley 1996). 
Detections of lumpiness in the velocity distribution of halo stars 
are becoming increasingly convincing, and the recent discovery
of the Sagittarius dwarf satellite galaxy 
(Ibata, Gilmore \& Irwin 1994) is a dramatic confirmation that
accretion and merging continue to affect the Galaxy.

There have been a number of recent studies of the accretion and
disruption of satellite galaxies (Quinn, 
Hernquist \& Fullagar 1993; Oh, Lin \& Aarseth 1995; 
Johnston, Spergel \& Hernquist 1995;  Vel\'azquez \& White
1995, 1999; Sellwood, Nelson \& Tremaine 1998). Much of this work
has been limited to objects which remain mostly in
the outer parts of the Galaxy, which may be  
well represented by a spherical potential plus a small perturbation due
to the disk (Johnston, Hernquist \& Bolte 1996; Kroupa 1997; 
Klessen \& Kroupa 1998). In this situation simple 
analytic descriptions of the disruption process, 
of the properties of the debris, etc. are possible (Johnston 1998).
However, it is questionable whether such descriptions can be applied 
to most of the regions probed by 
past or current surveys of the halo, which are quite local:
in this case the influence of the disk cannot be disregarded
or treated as a small perturbation.

Since formation models for  the Galaxy should address the
broader cosmological setting, 
we are naturally led to ask what should be the signatures of 
the different accretion events that our Galaxy may have suffered 
through its lifetime. 
Should this merging history be observable 
in star counts, kinematic or abundance surveys of the Galaxy? 
How prominent should such substructures be? How long do they survive, or
equivalently, how well-mixed today are the stars which 
made up these progenitors? 
What can we say about the properties
of the accreted satellites from observations of the present stellar
distribution?  
Our own Galaxy has a very important role in 
constraining galaxy formation models, 
because  we have access to 6-D information which is available for no
other system. Observable structure 
which could strongly constrain the history of the formation
of galaxies is just at hand.

This paper will try to answer some of the questions just posed.
We focus on the growth of the stellar halo of the Galaxy
by disruption of satellite galaxies. We have run numerical
simulations of this process, and have studied the properties
of the debris after many orbits, long after the disruption has taken
place. We analyse how the debris phase-mixes by following the growth
of its entropy and the variations of the volume it fills 
in coordinate space. We also study the evolution of its kinematical 
properties.  In order to model  the 
characteristic properties of the disrupted system, such as its size, 
density and velocity dispersion, we develop a simple analytic prescription 
based on a linearized Lagrangian treatment of its evolution in
action-angle variables.
We apply our results to derive the observable properties of an
accreted halo in the solar neighbourhood. We also analyse the 
clump of halo stars detected near the NGP by 
Majewski et al. (1994), and obtain an order of magnitude estimate
for the initial properties of the progenitor system.
 
Our paper is organized as follows. Section 2 presents our numerical
simulations. In Section 3 we analyse the characteristics of the debris
in these models, and in Section 4 we develop an analytic formalism to 
understand their properties. We apply this formalism to describe the 
characteristics of an accreted halo in this same section. 
In Section 5 we compare our modelling with the 
observations of Majewski et al. (1994). We leave for
the last section the discussion of the results, their validity, and the
potential of our approach for understanding the formation of our Galaxy.

\section{The simulations}

To study the disruption of a satellite galaxy of the Milky Way, we
carry out N-body simulations in which the Galaxy is represented by a 
fixed, rigid potential and the satellite by a collection of
particles. The self-gravity of the satellite is modelled by 
a monopole term as in White (1983) and Zaritsky \& White (1988).
 
\subsection{Model}
The Galactic potential is represented by two components:
a disk described by a Miyamoto-Nagai (1975) potential,
\begin{equation}
\label{eq:disk}
\Phi_{\rm disk} = - \frac{G M_{\rm disk}}{\sqrt{R^2 + (a + 
\sqrt{z^2 + b^2})^2}},
\end{equation}
where $M_{\rm disk} = 10^{11}\, {\rm M_{\odot}}$, $a = 6.5\, {\rm kpc}$, 
$b = 0.26 \,{\rm kpc}$, and a dark halo with a logarithmic potential,
\begin{equation}
\label{eq:halo}
\Phi_{\rm halo} = v^2_{\rm halo} \ln (r^2 + d^2),
\end{equation}
with $d = 12 \,{\rm kpc}$ and $v_{\rm halo} = 131.5 \, {\rm km\, s^{-1}}$.
This choice of the parameters gives a circular velocity at the solar
radius of \mbox{$210 \,{\rm km\,  s^{-1}}$}, 
and of \mbox{$200\, {\rm km \, s^{-1}}$} at $\sim 100$ kpc.

We have taken two different initial phase-space density distributions
for our satellites: $i$) two spherically symmetric Gaussian distributions in 
configuration and velocity space of 1 kpc (5 kpc) width and 
 $5 - 25$ \kms (20 \kms) velocity dispersion, corresponding to 
masses of $\sim 5.9 \times 10^7 - 1.5 \times 10^9 \sm$ 
($4.7 \times 10^9 \sm$); and $ii$) 
a Plummer profile (1911)
\begin{equation}
\label{eq:density_sat}
\rho(r) =  \frac{\rho_0}{(r^2 + r^2_0)^{5/2}},
\end{equation}
with $\rho_0 = 3 M/4 \pi r_0^3$, $M$ being the initial mass of 
the satellite and $r_0$ its scale length. In this second case, 
the distribution of
initial velocities is generated in a self-consistent way with 
the density profile. For the characteristic parameters we chose
$M = 10^{7} - 10^9 \, {\rm M_{\odot}}$ and  
$r_0 = 0.53 - 3.0 \, {\rm kpc}$, giving a 
one-dimensional internal velocity dispersion 
$\sigma_{1D} = 2.9 - 11.3 \, {\rm km \, s^{-1}}$.

The force on particle $i$ due to the self-gravity of the satellite 
is represented by 
\begin{equation}
\label{eq:self_grav}
{\mathbf F}({\mathbf x}_{i}) = - \frac{G M_{\rm in}}{(r_{i}^2 + 
\epsilon^2)^{3/2}} {\mathbf r}_i, 
\end{equation}
where $M_{\rm in}$ is the mass of the satellite inside 
$r_i = |{\mathbf x}_i - {\mathbf x}_c|$, 
${\mathbf x}_c$ being the position of the expansion centre 
defined by a test particle with the same orbital properties 
as those of the satellite. 
The value for the softening $\epsilon$ is $ 0.25 \,r_0$.
The approximation for the self-gravity 
of the satellite may not be very accurate during the disruption 
process, where tidal forces are strong 
and elongations in the
bound parts of the satellite are expected. However, 
because we are interested in what happens after many perigalactic
passages, well after the satellite has been tidally destroyed, 
our conclusions on the whole process are unaffected by details 
of the disruption process. 

In total we ran sixteen different simulations, six of which we analyse 
and describe in full detail in Section 3. Some of the
remaining simulations are used in Section 4 for comparison with the
analytic predictions and the rest are briefly mentioned in the discussion.
The characteristic properties of our six principal simulations are
summarized in Table~1. They differ only in their orbital
parameters and all initially have a Plummer profile and  a mass of 
$10^7 \sm$. We have imposed the restriction 
that the orbits pass close to the solar 
circle in order to be able to compare the results of the experiments 
with the known properties of the local stellar halo. In all cases 
the satellite was represented by $10^5$ particles of equal mass.
\begin{figure}
\label{fig1}
\center{\psfig{figure=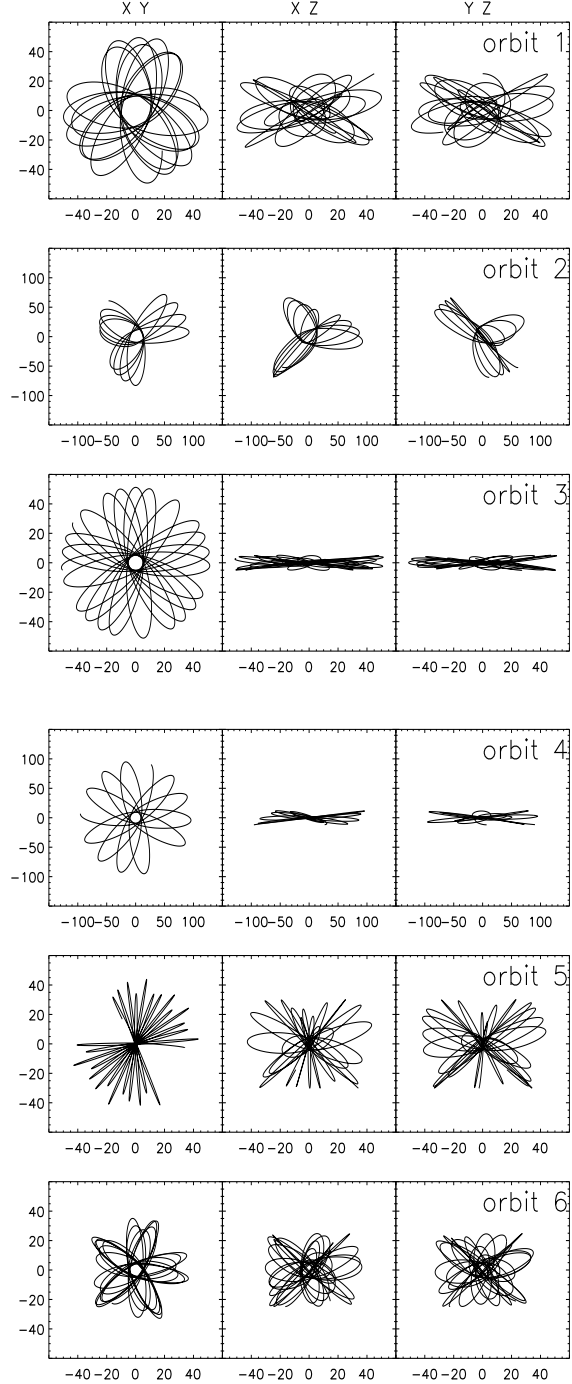,height=18.5cm,width=7.48cm}}
\caption[]{Projections of the orbits of the satellite on different 
orthogonal planes, where XY coincides with the plane of the Galaxy. All 
distances are in $\rm kpc$.}
\end{figure}
\begin{table}
\caption{0rbital parameters for the different experiments.}
\begin{tabular}{ccccc}
\hline
Experiment &  pericentre & apocentre & $z_{\rm max}$ & period  \\
 & (kpc) & (kpc) & (kpc) &(Gyr) \\
\hline
1 & 10.9 & 51.5 & 25.0 & 0.69 \\
2 & 13.5 & 93.1 & 69.1 & 1.23 \\
3 & 5.0  & 51.5 & 5.1 & 0.64  \\
4 & 9.2 & 96.5 & 12.0 & 1.24 \\
5 & 0.5 & 45.5 & 30.1 & 0.56 \\
6 & 6.0 & 37.0 & 24.8 & 0.48 \\
\hline
\end{tabular}
\end{table}

In Figure~1 we show projections of orbits 1--6 in three orthogonal
planes,
where XY always coincides with the plane of the Galaxy. Notice that
the plane of motion of a test particle on these orbits changes 
orientation substantially showing that the non-sphericity  induced 
by the disk significantly affects the motion of the satellite.

While orbiting the Galaxy, the satellite loses all of its mass.
As expected, the most dramatic effects take place during  
pericentric passages. The satellites do not survive very long, being 
disrupted completely after 3 passages. 
This means that for our experiments, for any relatively low density 
satellite on an orbit which plunges deeply into the Galaxy with a 
period of 1 Gyr or less, the disruption itself occupies
only a relatively small part of the available evolution time.

\section{Properties of the debris: Simulations}

\subsection{Entropy as a measure of the phase-mixing}

The state of a collisionless system is completely specified by its
distribution function $f({\mathbf x},{\mathbf v}, t)$. In making actual
measurements, it is often more useful to work with the 
coarse-grained distribution
function $\langle \, \!f\!\, \rangle $, which is the average of 
$f$ over small cells in phase-space.
An interesting property of 
the coarse-grained distribution function is that it can yield information
about the degree of mixing of the system
(Tremaine, H\'enon \& Lynden-Bell 1986; Binney and Tremaine 1987). 

In statistical mechanics the entropy is defined as
\begin{equation}
\label{eq:def_entropy}
S = -\int d^3x ~d^3v ~ f({\mathbf x},{\mathbf v},t) 
\ln f({\mathbf x},{\mathbf v},t).
\end{equation}\noindent
Since the coarse-grained distribution function decreases 
as the system evolves towards a well-mixed state, 
an entropy calculated using $\langle \, \!f\!\, \rangle $ will increase, 
whereas one calculated using $f$ will remain constant, a consequence of 
the collisionless Boltzmann equation: $Df/Dt = 0$. 
We therefore quantify the mixing state of
the debris by calculating its coarse-grained entropy as a function of 
time. We represent the coarse-grained distribution function by taking
a partition in the 6-dimensional phase-space and counting how
many particles fall in each 6-D box. Naturally the size chosen for
the partition and the discreteness of the simulations  
will affect the result. We can quantify the expected discreteness noise
in the following way.
\begin{figure*}
\label{fig2}
\flushleft{\psfig{figure=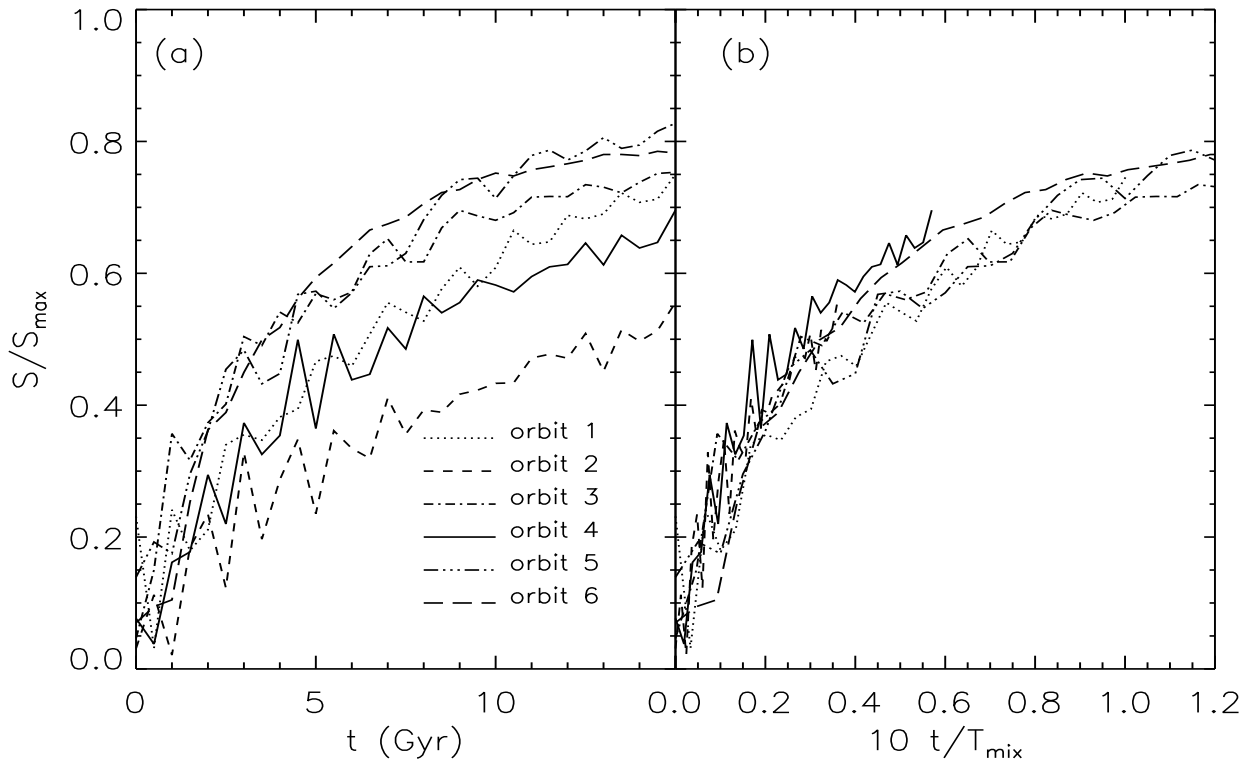,angle=90,height=10cm,width=16.4cm}}
\caption[]{Evolution of the entropy of the system for the different 
experiments, as a function of time in (a), and scaled with the mixing
time-scale in (b). The error in the scaled 
entropy is of the order of 0.06.}
\end{figure*}
The uncertainty in the entropy can be attributed to fluctuations in 
the number counts, which we can estimate as Poissonian, 
$\propto \sqrt{N_i}$ in each occupied cell. Therefore, the 
uncertainty in the entropy in each cell is
\[
\Delta S_i \approx \frac{\Delta N_i}{N} \left(1 + \ln \frac{N_i}{N}\right)
\approx \frac{\sqrt{N_i}}{N} \ln N
\]
for $N \gg 1$. The total uncertainty is thus
\begin{equation}
\label{eq:error_entr}
\Delta S \approx \frac{\ln N}{\sqrt{N}}
\end{equation}
which, for experiments with $10^5$ particles is $0.04$.
In order to have a normalized
measure of the mixing properties of the debris, we
also computed the entropy of points equidistant in time along the  
corresponding orbit. After a very
long integration, the orbit will fill the available region in 
phase-space, whose shape and size are
determined by its integrals of motion. In this way, by comparing
the entropy calculated for the debris with the `entropy of the orbit',
we have a measure of how well mixed 
the debris is. We plot this `normalized' entropy in \mbox {Figure~2(a)} as 
a function of time.
Note that the orbits which have the shortest periods show the most 
advanced state of mixing, but that this 
is not complete after a Hubble time. 

The degree of mixing 
basically depends on the range of orbital frequencies in the
satellite, essentially as $(\Delta \nu) ^{-1}$ (Merritt 1999). 
This means, for example, that a small satellite will disperse much
more slowly than a larger one on the same orbit. On the other hand
a satellite set close to a resonance will mix on a much longer
time scale. One can also imagine that if there are fewer isolating
integrals than degrees of freedom so that chaos might develop, a
satellite located initially in a chaotic region will have a 
large spread $\Delta \nu$ because of the extreme
sensitivity to the initial conditions. Therefore the mixing
timescale (no longer a {\it phase}-mixing timescale) will be very
short, since the neighbouring orbits diverge exponentially, 
instead of like power-laws.
If indeed the mixing rate is set by the spread in the orbital
frequencies $\nu$ of the satellite, by normalising the time variable with
this timescale we should be able to derive a unique curve for the 
entropy evolution $S = S_{\rm max} f(t/T_{\rm mix})$. 

In what follows we shall assume that the behaviour of the system
is regular as seems to be the case for our experiments.
Let us recall that any regular motion can be expressed as a 
Fourier series in three basic frequencies (Binney \& Spergel 1984, 
Carpintero \& Aguilar 1998). The motion is therefore a linear 
superposition of waves of the basic frequencies with different amplitudes.
Terms in this expansion which have the
largest amplitude will be the dominant terms and may be used to define
three independent (basic) frequencies.
By performing a spectral dynamics analysis 
as outlined by Carpintero \& Aguilar (1998) for ten randomly selected
particles in our satellites in each experiment, we compute the
frequencies associated with the largest amplitude terms in the $x$-  (or
$y$, since the problem is axisymmetric) and $z$-motions, and their
dispersion around the mean. We then define  
\begin{equation}
T_{\rm mix}^{-1} = min\{\sigma(\nu_{x}^{(1)}),\sigma(\nu_{x}^{(2)}), 
	\sigma(\nu_{z}^{(1)}),\sigma(\nu_{z}^{(2)})\},
\end{equation}
The curves obtained by scaling time with $T_{\rm mix}$ are shown 
in \mbox{Figure~2(b)} and they
can be well fitted with the function
\begin{equation}
\label{eq:entr_fit}
\frac{S}{S_{\rm max}} = 0.78 - 0.69 \,\exp(-27.03 \,\frac{t}{T_{\rm mix}}).
\end{equation}
The good fit and small dispersion confirms that mixing is governed
primarily by the spread in frequency.

\subsection{Configuration space properties}

To analyse the spatial properties of the debris several Gyr
after disruption, we have plotted smoothed isodensity 
surfaces and calculated different characteristic densities. In Figures~3 
and 4 we show the density surface at approximately $10^{-6}$ times the 
initial density of the satellite. This encompasses most of the 
satellite's mass.
\begin{figure}
\label{fig3}
\center{\psfig{figure=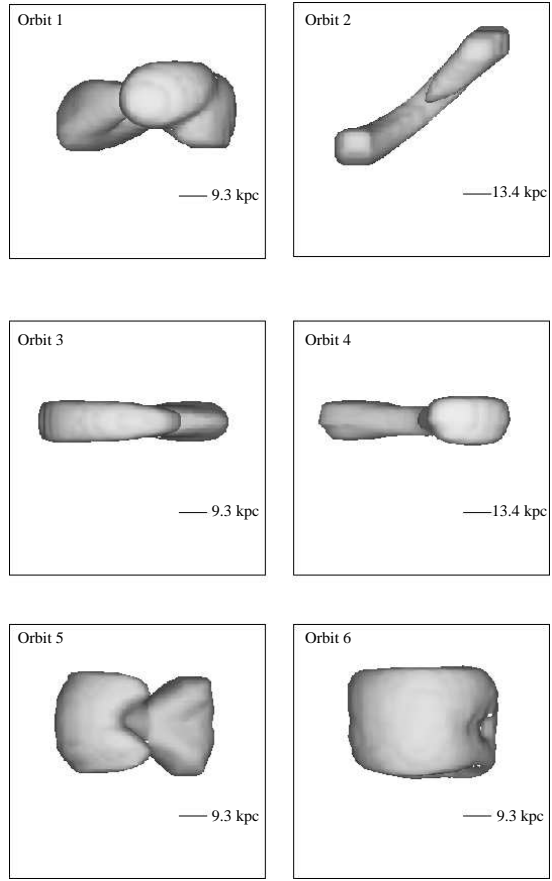,height=11.7cm}}
\caption[]{Isodensity surface of $10^{-6} \rho_0$ after 14 Gyr, seen from
the Galactic plane, for the different experiments. }
\end{figure}
\begin{figure}
\label{fig4}
\center{\psfig{figure=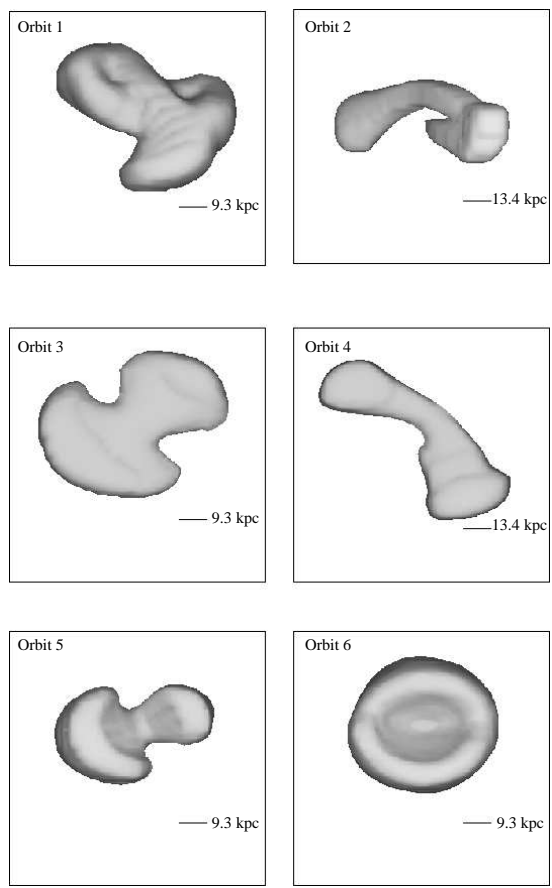,height=11.7cm}}
\caption[]{Isodensity surface of $10^{-6} \rho_0$ after 14 Gyr, seen from
the Galactic pole, for the different experiments. }
\end{figure}
This density surface practically does not
change over the last 2 Gyr for experiments 3, 5, 6, showing 
that the system has reached a stage where it fills 
most of its available 3-D coordinate space. 
The shape of this isodensity surface also gives a measure of how advanced 
the disruption is. The form of the accessible \mbox{3-D} configuration 
volume is basically 
a torus, defined by the apocentre, pericentre
and the inclination of the orbit. In Figures~3 and 4 we clearly see that
shape for experiment 6. Experiments 3 and 5 are in an intermediate state
and still need to fill part of their tori. In the opposite limit,
experiment 2 has filled only a small fraction of its available volume.
All this
is consistent with what was found using the entropy in
the previous subsection. The characteristic extent of the debris is 
much larger than the initial size of the satellite. Moreover, 
debris with these properties may well span a very large solid angle
on the sky, and so be poorly described
as a stream in coordinate space. This is the principal difference between
our own experiments and those
in which the Galaxy is represented by a spherical potential. In the latter
the plane of motion of the satellite has a fixed orientation, and therefore
all the particles have to remain fairly close to this plane, 
naturally giving a stream-like configuration. Late accretion
events in the outer halo of the Galaxy will plausibly have this 
characteristic, as shown in Johnston et al. (1996) and Johnston (1998). 
However, similar behaviour
should not be expected in the solar neighbourhood, or even as far as 
\mbox{10--15 kpc} from the galactic centre since at such radii no strong 
correlations are left in the spatial distribution of satellite particles. 
Any method which attempts to find moving groups 
purely by counting stars will probably fail in this regime.

In Table~2, we present a summary of characteristic densities 
at different times 
which were calculated by counting particles within spheres of
$0.5 \, {\rm kpc}$ radii. The maximum
density is achieved at 
the pericentre of the orbit, though most of the mass 
is distributed closer to the apocentre. 
In all cases the maximum density is between three and four orders 
of magnitude lower than the initial density of the satellite, 
and the mean density of the debris is
between four and five orders of magnitude lower. 
These values give another estimate of the degree of mixing
of the debris. Note that, in accordance with the entropy computation, 
experiment 6 has the smallest characteristic densities,  
meaning that it has reached a rather evolved state,
whereas experiment 2 has high densities in comparison 
to the rest. 
The maximum density in all of the experiments is roughly
comparable (similar or an order of magnitude lower) to
the local density of the Milky Way's stellar halo, 
though the sizes of regions where this
density is reached get fairly small, a few ${\rm kpc}^3$, as the evolution
proceeds.
\begin{table}
\caption{Characteristic densities for the different experiments.}
\begin{tabular}{cccc}
\hline
Experiment & time  & $\rho_{\rm mean}$ & $\rho_{\rm max}$ \\
 & Gyr & $ 10^{2}{\rm M_{\odot}~kpc^{-3}} $ & 
$10^{2}{\rm M_{\odot}~kpc^{-3}}$ \\
\hline
\hline
1 & 5.0 & 67.0 & 886.2 \\
  & 10.0 & 14.6 & 223.5 \\
  & 12.5 & 7.0 & 152.8 \\
  & 15.0 & 6.8 & 181.4\\
\hline 
2 & 5.0 & 84.7 & 857.5 \\
  & 10.0 & 26.5 & 376.2\\
  & 12.5 & 11.5 & 202.4 \\
  & 15.0 & 9.7 & 288.4 \\
\hline
3 & 5.0 & 41.5 & 437.4 \\
  & 10.0 & 8.9 & 72.6 \\
  & 12.5 & 8.7 & 181.4 \\
  & 15.0 & 6.9 & 177.6 \\
\hline
4 & 5.0 & 40.8 & 446.9 \\
  & 10.0 & 5.9 & 99.3 \\
  & 12.5 & 5.7 & 171.9 \\
  & 15.0 & 5.1 & 156.6 \\
\hline
5 & 5.0 & 36.4 & 996.9 \\
  & 10.0 & 10.9 & 210.1 \\
  & 12.5 & 6.1 & 183.3 \\
  & 15.0 & 5.7 & 213.9 \\
\hline
6 & 5.0 & 13.8 & 403.0 \\
  & 10.0 & 4.3 & 82.1 \\
  & 12.5 & 4.3 & 95.5 \\
  & 15.0 & 3.4 & 63.0 \\
\hline
\end{tabular}
\end{table}

\subsection{Velocity space properties}

Let us now focus on the characteristics of the debris
in velocity space. We divided the 3-D coordinate space into 
boxes and analysed the kinematical
properties of the particles inside each box. Figure~5 shows an example.
The scatter diagrams indicate that there is a strong correlation
between the different components of the velocity vector inside any 
given box. Notice also the large velocity range in each component
when close to the Galactic centre. This
shows that the debris can appear kinematically hot. As we shall see this 
results from a combination of multiple streams within a
given box (clearly visible in Figure~5) and of  
strong gradients along each stream.
At a given point on a particular stream
the dispersions are usually very small.

\begin{figure*}
\label{fig5}
\center{\psfig{figure=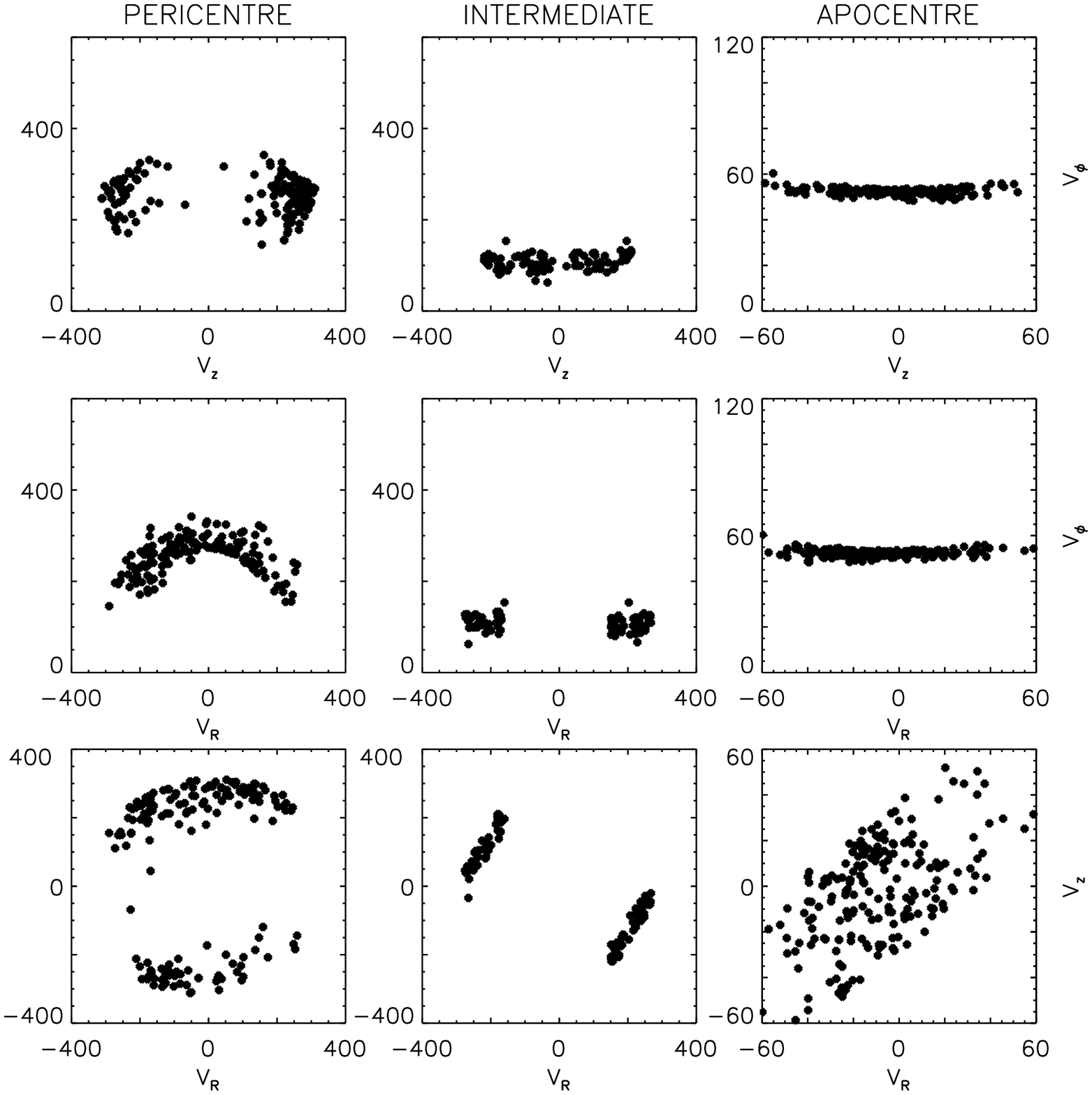,height=12cm,width=11.6cm}}
\caption[]{Scatter plots of the different velocity components for stars
in boxes of $\sim$ 3 kpc on a side at different locations for
experiment 6 at 13.5 Gyr. 
Similar characteristics are observed in all our experiments.} 
\end{figure*}

\section{Properties of the debris: Analytical approach}

In this section we will develop an analytic formalism to 
understand and describe the spatial and kinematical properties of the 
stream.
Let us recall that because the disruption of the satellite
occurs very early in its history, the stars that were once
part of it behave as test particles in a rigid potential for most of the
evolution. One of the distinguishing properties of this ensemble of
particles is that it initially had a very high density in phase-space,
and by virtue of Liouville's theorem, this is true at all times.
At late times, however, this is no longer reflected by a strong
concentration in configuration space. This evolution can be 
understood in terms of a mapping from the initial configuration to the 
final configuration, which we will describe by using the adiabatic 
invariants, namely the actions.

\subsection{Action-Angle variables and Liouville's theorem }
\label{sec:general}  

Let $H = H({\mathbf{q}},{\mathbf{p}})$ be the (time-independent) 
Hamiltonian of the problem
and $({\mathbf{q}},{\mathbf{p}})$ a set of canonical coordinates. 
We wish to transform the initial set $({\mathbf{q}},{\mathbf{p}})$ to
one in which the evolution of the system is simpler, for
example, where all the momenta $P_i$ are constant. To meet this last
condition, it is sufficient to require that the new Hamiltonian be
independent of the new coordinates $Q_i$: $H = H({\bf P}) = E$. 
The equations of
motion then become
\[
\dot{Q_i} = \nu_i, \qquad \dot{P_i} = 0,\] \noindent
with solutions
\[
 Q_i = Q_i^0 + \nu_i t, \qquad P_i = P_i^0.
\]
The generating function that produces this transformation is known
 as Hamilton's
Characteristic function $W({\mathbf{q}},{\mathbf{P}})$, and satisfies the 
Hamilton-Jacobi partial differential equation:
\[
H(q, \frac{\partial W}{\partial q}) = E.
\]
The solution to this equation involves $N$ constants of integration 
$\alpha_i$ (including $E$) for a system with $2N$ degrees of freedom. 
Therefore, the new momenta {\bf P} may be chosen as functions of 
these $N$ constants of integration.
A particularly simple situation occurs if the potential is 
separable in the original  coordinate set
$({\mathbf{q}},{\mathbf{p}})$. The characteristic function may then be
expressed as $W = \sum_i W_i(q_i,\alpha_1...\alpha_N)$, and
the Hamilton-Jacobi equation breaks up
into a system of $N$ independent equations of the form:
\[
H_i\left(q_i, \frac{\partial W_i}{\partial q_i}, \alpha_1...\alpha_N\right) = 
\alpha_i,
\]
each of which involves only one coordinate and the partial derivative 
of $W_i$ with respect to that coordinate. The transformation relations 
between the original and new sets of variables are
\[
p_i = \frac{\partial W}{\partial q_i}, \qquad 
Q_i = \frac{\partial W}{\partial P_i},
\]
and each component of the characteristic function is given by
\begin{equation}
\label{eq:W}
W_i(q_i,\alpha_1..\alpha_N) = \int dq_i' \, 
p_i(q_i',\alpha_1..\alpha_N).
\end{equation}
(For more details, e.g. Goldstein 1953).

The actions and angle variables are a set of coordinates that 
describe simply the evolution of a system of particles. They are
particularly useful in problems where the motion is periodic. The actions
are functions of the constants $\alpha_i$ and are defined for
a set of coordinates $({\mathbf{q}},{\mathbf{p}})$ as
\begin{equation}
\label{eq:defJphi}
J_i= \frac{1}{2 \pi} \oint dq_i\, p_i, 
\end{equation}
and their conjugate coordinates, the angles, are
\begin{equation}
\phi_i= \frac{\partial W}{\partial J_i}.
\end{equation}
The evolution of the dynamical system thus becomes:
\begin{eqnarray}
\label{eq:evol}
\phi_i\!\!\!& = &\!\!\! \phi_i^0 + \Omega_i({\mathbf{J}})\, t, \nonumber\\
J_i \!\!\!& = &\!\!\! J_i^0 = constant.
\end{eqnarray}

\subsubsection{The evolution of the distribution function}

Let us assume that the initial distribution function of the ensemble 
of particles is a multivariate Gaussian in configuration and velocity space
\[
f({\bf x}, {\bf v},t^0) = f_0 \exp{\left[-\sum_{i=1}^3 \frac{(x_i-
\bar{x}_i^0)^2}
{2 \sigma_x^2} 
\right]}
\exp{\left[-\sum_{j=1}^3\frac{(v_j-\bar{v}_j^0)^2}{2 \sigma_v^2}
\right]},
\]
which we can also express using matrices as
\begin{equation}
\label{eq:arg_ini}
f({\bf x}, {\bf v},t^0) = f_0 \exp{\left[-\frac{1}{2}
{{\bf \Delta}_\varpi^0}^{\dagger} {\bf \sigma}_\varpi^0 {\bf \Delta}_\varpi^0
\right]}.
\end{equation}
Here $t^0$ denotes the initial time.
${\bf \Delta}_\varpi^0$ is a 6-dimensional vector, 
with three spatial and three velocity components, and 
${{\bf \Delta}_\varpi^0}^\dagger$ is obtained by 
transposing ${\bf \Delta}_\varpi^0$. 
Explicitly
${\Delta_\varpi^0}_i = x_i - \bar{x}_i^0$ for $i=1..3$ and 
${\Delta_\varpi^0}_{i} = v_j - \bar{v}_j^0$ for $i=j+3=4..6$
in a Cartesian coordinate system.  
The matrix ${\bf \sigma}_\varpi^0$ is diagonal with ${\sigma_\varpi^0}_{ii} = 
 1/\sigma_x^2$ 
for i=1..3, 
and ${\sigma_\varpi^0}_{ii} = 1/ \sigma_v^2$ for i=4..6. 
As we shall see the matrix
formulation is particularly useful to study the evolution of the 
distribution of particles of the system.

At the initial time, we perform a coordinate change from Cartesian
to action-angle variables. Since the
particles are initially strongly  clustered in phase-space, a linearized
transformation 
can be used to obtain the distribution function of the whole system in the 
(${\bf \phi},\,{\bf J}$) variables.
We express this coordinate transformation as
\begin{equation}
{\bf \Delta}_\varpi^0 = {\bf T}^0 {\bf \Delta}_{w}^0, \qquad \mbox{with} 
\qquad T_{ij}^0 = \frac{\partial \varpi_i}{\partial w_j}
\bigg\vert_{\bar{{\bf x}}^0, {\bar{\bf v}}^0},
\end{equation}
where ${\bf \varpi} = (\bf{x},\,\bf{v})$, $w = (\bf{\phi},\,\bf{J})$ 
and the 
elements of matrix ${\bf T}^0$ are 
evaluated at the central point of the system, around which the expansion is
performed. 
By substituting this in Eq.~(\ref{eq:arg_ini}), and by defining
$
{\bf \sigma}_w^0 = {{\bf T}^0}^{\dagger} {\bf \sigma}_\varpi^0 {\bf T}^0$ 
the  distribution function in action-angle coordinates becomes
\begin{equation}
\label{eq:arg_ini_aa}
f({\bf \phi}, {\bf J},t^0) = f_0 \exp{
\left[-\frac{1}{2}{{\bf \Delta}_w^0}^{\dagger}{\bf \sigma}_w^0 
{\bf \Delta}_w^0\right]},
\end{equation}
that is, it is also a 
multivariate Gaussian, but with dispersions now given by $\sigma_w^0$.

The deviation of any individual orbit from the mean orbit, defined by the
centre of mass or the central particle of the system, 
$\Delta_{w_i} = w_i - \bar{w_i}(t)$ may in turn
be expressed in terms of the initial action-angle variables as
\begin{equation} 
\label{eq:j}
J_i - \bar{J}_i = J_i^0 - \bar{J}_i^0,
\end{equation}
and
\begin{equation}
\label{eq:omega'}
\phi_i - \bar{\phi}_i(t) = \phi_i^0 - \bar{\phi}_i^0  +  
\frac{\partial \Omega_i}{\partial J_k}\bigg\vert_{\bar{\bf J}}
 (J_k - \bar{J}_k)\, t,
\end{equation}
where we expanded the difference in the frequencies to first order in 
$J_k - \bar{J}_k$. 
Eqs.~(\ref{eq:j}) and (\ref{eq:omega'}) can also be written as
\begin{equation}
{\bf \Delta}_w(t) = {\bf \Theta}^{-1}(t) {\bf \Delta}_w^0, 
\end{equation}
where ${\bf \Theta}(t)$ is the blockmatrix:
\begin{equation}
\label{eq:matrix_th}
{\bf \Theta}(t) = \left[\begin{array}{cc} 
 {\bf{\cal I}_3} & - {\bf \Omega'} t \\
 {\bf 0} & {\bf{\cal I}_3} \end{array}\right].
\end{equation}   
${\bf{\cal I}_3}$ here is the identity matrix in 3-D, and 
${\bf  \Omega'}$ represents a $3\times3$ matrix whose elements are 
$\partial \Omega_i/\partial J_j$. 
The distribution function in action-angle 
space in the neighbourhood of the central particle at any point
of its orbit $(\bar{{\bf \phi}}(t), \bar{\bf J})$ is then
\begin{equation}
\label{eq:df-aat}
f({\mathbf \phi}, {\bf J},t) = f_0 \exp{
\left[-\frac{1}{2}{\bf \Delta}_w^{\dagger}(t) 
{\bf \sigma}_w(t) {\bf \Delta}_w(t)\right]},
\end{equation}
with ${\bf \Delta}_w(t) = ({\bf \phi} - \bar{\bf \phi}(t), {\bf J} - 
\bar{\bf J})$
and 
\begin{equation}
{\bf \sigma}_w(t)= {{\bf\Theta}(t)}^{\dagger} {\bf \sigma}_w^0{\bf \Theta}(t),
\end{equation}
or in terms of the original coordinates
$
{\bf \sigma}_w(t)= ({\bf T}^0 {\bf \Theta}(t))^{\dagger} 
{\bf \sigma}_\varpi^0 ({\bf T}^0 {\bf \Theta}(t)).
$

\vspace{0.3cm}
{\it Example: 1-D Case.}
To understand more clearly what the distribution function in 
Eq.~(\ref{eq:df-aat}) tells us
with respect to the evolution of the system, we 
consider the 1-D case. The initial distribution function becomes:
\[f(\phi, J,t^0)= f_0 \exp\left[-\frac{(\phi-\bar{\phi}^0)^2}
{2 \sigma_\phi^2}
 - \frac{(J-\bar{J})^2}{2 \sigma_J^2} - 
(\phi-\bar{\phi}^0)(J-\bar{J}) C_{\phi J}\right], \]
where $C_{\phi J}$ denotes the initial correlation\footnote{$C_{\phi J}$
 is not the correlation 
coefficient, usually denoted as $\rho$. They are related through 
$\rho = \frac{- C_{\phi J} \sigma_\phi^2 \sigma_J^2}
{1 - C_{\phi J}\sigma_\phi^2 \sigma_J^2}$.}  between $\phi$ and $J$.
After considering the time evolution of the system 
(as in Eq.~(\ref{eq:omega'})) we find
\[
f(\phi, J,t)  =  f_0 \exp \left[-\frac{(\phi-\bar{\phi}(t))^2}
{2 \sigma_\phi^2} -
(J-\bar{J})^2 \left(\frac{1}{2 \sigma_J^2} + \frac{{\Omega'}^2 t^2}
{2 \sigma_\phi^2}\right) - (\phi-\bar{\phi}(t))(J-\bar{J}) 
\left(C_{\phi, J} + \frac{{\Omega'} t}{\sigma_\phi^2}\right)\right],\]
where $\Omega' = d \Omega/d J$.
This means that the dispersion in the $J$-direction effectively decreases
in time and the covariance between $\phi$ and $J$ increases with time. The
system becomes an elongated ellipsoid in phase-space as time passes
by as a consequence of 
the conservation of the local phase-space density. 
This evolution is illustrated in Figure~6. 

\begin{figure*}
\label{fig:aa}
\center{\psfig{figure=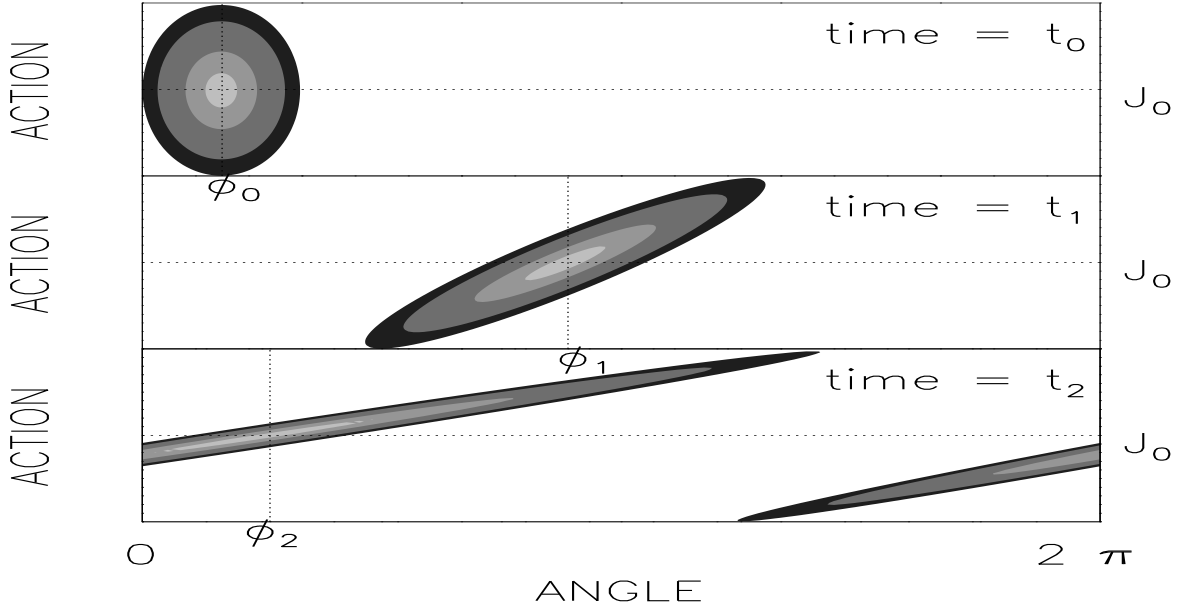,height=8cm,width=15.5cm}}
\caption{1-D graphical interpretation of Liouville's theorem and
the evolution of the system in phase space. The system is initially a
Gaussian in action-angle space, with no correlations between $\phi$ and $J$.
As time passes by, the system evolves into an ellipsoidal configuration,
with principal axes that are no longer aligned with the action or the angle
directions. After a some time, the system wraps around in the angles,
giving rise to phase-mixing: at the same phase we observe more
than one stream, each with a small variance in the action 
due to the conservation of the area in phase-space.}
\end{figure*}

\subsubsection{The distribution function in observable coordinates}

To compute the characteristic
scales of a system that evolved from an initial clumpy configuration, such
as satellite debris, we have to relate
the dispersions in action-angle variables to dispersions
in a set of observable coordinates.
The transformation from the action-angle coordinate system 
to the observable $({\bf x}, {\bf v})$ has to be 
performed locally since we generally cannot express in a simple way 
the global relation 
between the two sets of variables. Because the system has expanded
so much along some directions in phase-space, 
the transformation from (${\bf \phi}$, ${\bf J}$) to 
$({\bf x}, {\bf v})$ has to
be done point to point along the orbit. This transformation
is given by the inverse of ${\bf T}$ at time $t$:
\begin{equation}
T^{-1}_{ij} = \frac{\partial w_i}{\partial \varpi_j}
\bigg\vert_{{\bf x}, {\bf v}}, \noindent
\end{equation}\noindent
where the derivatives are now evaluated at the particular point of the
orbit around which we wish to describe the system in 
$({\bf x}, {\bf v})$ coordinates.
In particular, if the expansion is performed 
around $(\bar{\bf \phi}(t), \bar{\bf J})$ then
\begin{equation}
{\bf \Delta}_{w}(t) = {\bf T}^{-1} {\bf \Delta}_\varpi(t),
\end{equation}
and the distribution function may be expressed in the region around 
$\bar{\bf \varpi} = (\bar{\bf x}, \bar{\bf v})$ as
\begin{equation}
\label{eq:df_qpt'}
f({\bf x}, {\bf v},t) = f_0 \exp{\left[-\frac{1}{2}
{\bf \Delta}_\varpi(t)^{\dagger} {\bf \sigma}_{\bf \varpi}(t) 
{\bf \Delta}_\varpi(t)\right]},
\end{equation}
with
\begin{equation}
{{\bf \Delta}_\varpi}_i(t) = \left\{ \begin{array}{cc} 
x_i - \bar{x}_i(t), &  i=1..3, \\
 v_j - \bar{v}_j(t), & i = j+3 =4..6, \end{array}\right. 
\end{equation}
and 
\begin{equation}
\label{eq:sigma_final}
{\bf \sigma}_{\bf \varpi}(t) = ({\bf T}^0 {\bf \Theta}(t) 
{\bf T}^{-1})^{\dagger} 
{\bf \sigma}_{\bf \varpi}^0 ({\bf T}^0 {\bf \Theta}(t) {\bf T}^{-1}).
\end{equation}
We find once more that, locally, the distribution function is a multivariate
Gaussian, where the variances and covariances depend on their initial
values, on the time evolution of the system and on the position along the
orbit where the system centre is located at time $t$. 

If we wish to describe the properties of a group of particles that are located
at a different point ${\bf \tilde w}$ than the central particle (i.e.
the expansion centre does not coincide with the satellite centre at time $t$) 
a slightly different approach must be followed. The region of interest is
then ${\bf \Delta}_{w}(t) =   w' -  \bar w(t) = 
(w' - {\tilde w}) +
(- \bar{w}(t) + {\tilde w} ) = {\bf \Delta}'_{w} + 
{\bf \tilde{D}}(t)$.
We replace this in Eq.~(\ref{eq:df-aat}) and write
\begin{equation} 
\label{eq:df-aat_n}
f({\bf \phi}, {\bf J},t) = f_0 \exp{
\left[-\frac{1}{2} \left({\bf \Delta}'_w - \tilde{\bf D}(t)\right)^{\dagger} 
{\bf  \sigma}_w(t) 
\left( {\bf \Delta}'_w - \tilde{\bf D}(t)\right) \right]},
\end{equation}
or equivalently
\begin{equation}
\label{eq:df-aat_n1}
f({\bf \phi}, {\bf J},t)  = f_0'(t) \exp
\left[-\frac{1}{2}{{\bf \Delta}'_w}^{\dagger} {\bf  \sigma}_w(t) 
{\bf \Delta}'_w - \frac{1}{2}
{{\bf \Delta}'_w}^{\dagger} {\bf \sigma}_w(t) \tilde{\bf D}(t) 
 - \frac{1}{2}
\tilde{\bf D}(t)^\dagger {\bf \sigma}_w(t) {\bf \Delta}'_w 
\right],
\end{equation}
where $f_0'(t) = f_0 \exp{[-1/2 \,\tilde{\bf D}(t)^{\dagger} 
{\bf \sigma}_w(t)\tilde{\bf D}(t)]}$. 
We may now express
\mbox{${\bf \Delta}'_{w} = {\bf T'}^{-1} {\bf \Delta}'_\varpi$}, 
since the transformation is local again. The distribution function becomes
\begin{equation}
\label{eq:df-qpt_1}
f({\bf x'}, {\bf v'},t) = \tilde{f}_0(t) \exp{
\left[-\frac{1}{2}({\bf \Delta}'_\varpi - {\bf \delta}(t))^{\dagger}  
\sigma_{\bf \varpi'}(t) 
({\bf \Delta}'_\varpi - {\bf \delta}(t)) \right]},
\end{equation}
with
\begin{equation}
{\bf \delta}(t) = {\bf T}'\tilde{\bf D}(t), \qquad \qquad 
{\bf \sigma}_{\bf \varpi'}(t) = ({\bf T}'^{-1})^{\dagger}{\bf \sigma}_w(t)
{\bf T}'^{-1}, 
\end{equation}
and $\tilde{f}_0(t) = f_0'(t)\exp{[-1/2\,({\bf T}^{-1}{\bf \delta}(t))^\dagger 
{\bf \sigma}_w(t){\bf T}^{-1}{\bf \delta}(t)]}$. 
This means that the local distribution function is Gaussian centered
around ${\bf x_{m}} = {\bf {\tilde x}}
+ {\bf \delta}(t) $, which in general will not be very different from 
${\bf \tilde x}$,  with variances given by the
elements of ${\bf \sigma}_{\bf \varpi'}(t)$. Thus the same type of
behaviour as derived for the region around the system centre holds also
if far from it. 

The formalism here developed is completely general, but the
actions will not always be easy to compute.
As we mentioned  briefly in the beginning of this section, this depends
mainly  on whether the potential is separable in some set of coordinates.
We focus on the spherical case and a simple axisymmetric potential
in the next section
to show how this procedure can be
used to describe the characteristic scales of the debris. We refer the
reader to the Appendix for details of the computation.

\subsection{Spherical Potential}

\subsubsection{Analytic predictions}

For a spherical potential $\Phi(r)$, the Hamiltonian  is separable in 
spherical coordinates
and depends on the actions $J_\varphi$ and $J_\theta$ only through the
combination $J_\varphi + J_\theta = L$. This means that the problem can be
reduced to 2-D, and so we
may choose a system of coordinates which coincides with the plane of
motion of the satellite centre. The position of a particle 
is given by its angular ($\psi$) and radial ($r$) coordinates on that plane. 
Thus
\begin{eqnarray}
\label{eq:jr}
L \!\!\!& = &\!\!\! J_\psi = p_\psi, \nonumber \\
J_r \!\!\!& = &\!\!\! \frac{1}{\pi} \int_{r_1}^{r_2} dr \frac{1}{r}
\sqrt{2 (E - \Phi(r))\, r^2 - L^2}, 
\end{eqnarray}
where $ L$ is the total angular momentum of the particle, $E$ its energy 
and $r_1$ and $r_2$ are the turning points in the radial direction of motion. 
The frequencies of motion and their derivatives needed to
compute the matrix ${\bf \Theta}(t)$ and to obtain the
time evolution of the distribution function,   can be obtained by 
differentiating the  implicit function $g = g(E, L, J_r) \equiv 0$
defined by Eq.~(\ref{eq:jr}).

Let us assume that the variance matrix 
\footnote{Strictly speaking ${\bf \sigma}$ is the inverse of the covariance
matrix. However we will loosely refer to ${\bf \sigma}$ as the 
variance matrix.} in action-angle variables
is diagonal at $t=0$. This simplifies the algebraic computations
and, since we are only trying to calculate late-time behaviour, 
this assumption does not have a major influence on our results.
As shown in the previous section,  the evolution
of the system in action-angles is obtained through 
${\bf \sigma}_w(t) = {\bf \Theta}(t)^\dagger {\bf \sigma^0}_w {\bf \Theta}(t)$.
We find the properties of the debris in configuration and
velocity space by transforming the action-angle coordinates $w =
(\bf{\phi}, \bf{J})$ locally to
the separable $\bf{\omega} = (\bf{x}, \bf{p})$, and then by
transforming from $\bf{\omega} = (\bf{x}, \bf{p})$ to 
$\varpi = (\bf{x}, \bf{v})$. That is ${\bf \sigma_{\varpi}}(t) = 
{\bf T'}^\dagger {\bf \sigma}_w(t){\bf T'}$, 
with the $T' = T_{w \rightarrow {\bf \omega}} T_{p \rightarrow v}$.

The diagonalization of the variance matrix $\sigma_{\varpi}(t)$
yields the values of the dispersions along the principal axes
and their orientation. It can be shown that {\em two of the eigenvalues
increase with time}, whereas the other {\em two decrease with time}. This is
directly related to what happens in action-angle variables: as we
have shown for the 1-D case, the system becomes considerably elongated
along an axis which, after a very long time, is parallel
to the angle direction. For 2-D (\mbox{3-D}), the evolution in action-angles
can also be divided into two (three) independent motions 
(whether or not the Hamiltonian is separable), so that along each of
these directions this same effect can be observed. 
The directions of expansion and contraction 
are linear combinations of the four axes $({ \breve{\epsilon}_\psi}, 
\breve{\epsilon}_r, \breve{\epsilon}_{v_\psi}, \breve{\epsilon}_{v_r})$
and, generally, none is purely spatial or a pure velocity direction. 

To understand the properties of the debris in observable coordinates, we
will examine what  happens around a particular point
in configuration space. This is equivalent to 
studying the velocity part of the variance matrix: $\sigma_{\varpi}(v)$.
For example, by diagonalising the matrix $\sigma_{\varpi}(v)$ we obtain the 
principal axes of the velocity ellipsoid at the point  $\bar{\bf x}$.
Its eigenvalues are the roots of $\det[{\bf \sigma}_{\varpi}(v) -
\lambda {\bf{\cal I}}] = 0$. For $t \gg t_{\rm orb}$
\begin{eqnarray}
\lambda_1 \lambda_2 \!\!\!& = &\!\!\! t^4 \,
(\Omega'_{11} \Omega'_{33}-{\Omega'_{13}}^2)^2 r^2 \frac{p_r^2}{\Omega_r^2} 
\sigma_{11} \sigma_{33},
 \nonumber \\
\lambda_1 + \lambda_2 \!\!\!& = &\!\!\! t^2 r^2 
\left[ \sigma_{11} \left(\Omega'_{11} - \frac{\Omega'_{13}}{\Omega_r} 
\left(\Omega_\psi - \frac{L}{r^2}\right) \right)^2 
+ \sigma_{33} \left(\Omega'_{13} -
\frac{\Omega'_{33}}{\Omega_r} 
\left(\Omega_\psi - \frac{L}{r^2}\right)\right)^2 \right] \nonumber \\
&+&t^2 \left[ \sigma_{11}{\Omega'_{13}}^2 +
 \sigma_{33}{\Omega'_{33}}^2\right]\frac{p_r^2}{\Omega_r^2}, \nonumber
\end{eqnarray}
where the subindices 1 and 3 represent $\psi$ and $r$ respectively,
and $\sigma_{ii} = 1/\sigma_{\phi_i}^2$, the initial variance in the
angles. Since $\sigma(v_i) = \sqrt{1/\lambda_i}$ 
both directions in velocity space have decreasing 
dispersions on the average. 

So far we did not describe how the debris is spread
along the transverse direction to the plane of motion: 
$\breve{\epsilon}_\vartheta$ and
$\breve{\epsilon}_{v_\vartheta}$. This
is because we reduced the problem to 2-D in configuration space. However, the
problem is not really 2-dimensional since the system has a finite
width in the direction transverse to the plane of motion. Now
that we have understood the dynamics of the reduced problem, 
the generalization to \mbox{3-D} is straightforward.  
If the variance matrix initially is diagonal in action-angle variables, 
then the dispersions along $\phi_\vartheta$
and $J_\vartheta$ do not change  because the
frequency of motion in the transverse direction is zero.
Thus the velocity dispersion and width of the stream also 
remain unchanged in the direction perpendicular to the orbital plane.

By integrating Eq.~(\ref{eq:df_qpt'}) 
with respect to the velocities, 
we compute the  density at the point $\bar{\bf x}$
\begin{equation} 
\rho({\bf \bar{x}},t) = \int_{\Delta v_r}\int_{\Delta v_\varphi} 
\int_{\Delta v_\theta} dv_\theta \,dv_\varphi \,dv_r \, 
f({\bf \bar{x}}, {\bf v},t).
\end{equation}
For $t \gg t_{\rm orb}$, 
\begin{equation} 
\label{eq:rho_sph}
\rho({\bf \bar{x}},t)= \frac{(2 \pi)^{3/2} f_0 
\sigma_{\phi_3}}
{|\Omega'_{11} \Omega'_{33}-{\Omega'_{13}}^2|}
\left[\sqrt{\left(\frac{1}{\sigma_{\phi_1}^2}+
\frac{1}{\sigma_{\phi_2}^2}\right) \left(\frac{1}{\sigma_{J_1}^2} +
\frac{1}{\sigma_{J_2}^2}\right)}\right]^{-1}
\frac{\Omega_r L}{r^2 \sin\theta |p_r p_\theta|}
\frac{1}{t^2},
\end{equation}
where $\sigma_X$ is the initial dispersion in the quantity $X$.
This equation shows that the density at the central point of the 
system decreases, on the average, as $1/t^2$. 
It tends to be larger near pericentre since it depends on radius as $1/r^2$; 
moreover it diverges at the turning points of the orbit. 
Even though the system evolves smoothly in action-angle 
variables, when this behaviour is projected onto observable space, 
singularities arise associated with
the coordinate transformation. In action-angle variables the motion is 
unbounded, whereas in configuration space
the particle finds itself at a `wall'
near the turning points. This divergence shows up in the elements
of the transformation matrix  $T_{w \rightarrow \varpi}$ 
(Eq.~(\ref{eq_aptransf_el})), some of which tend to zero, while  
others diverge keeping the matrix non-singular.  
Because of the secular evolution of the dispersions,  
the intensity of the spikes will decrease with time. They are generally
stronger at the pericentre of the orbit than at the apocentre, because
of the $1/r^2$ dependence of the density. 

A direct consequence of the secular evolution is that the characteristic
sizes of the system, the width and length of the stream, will increase
linearly with time, reflecting the conservation of the full 6-D phase-space 
density. At the turning points one of these scales becomes extremely small. 
In Figure~7 we plot the predicted behaviour of the dispersions
along the principal axes of the velocity ellipsoid
as a function of time. 
We have chosen for the initial conditions a spherically symmetric 
Gaussian in configuration and velocity space. We follow 
the evolution of the variance matrix and, in particular, of the
velocity dispersions along the three principal axes at the positions of
the central particle. In all panels we can clearly see the periodic
behaviour associated with the orbital phase of the central particle, 
superposed on the secular
behaviour related to the general expansion of the system along the 
two directions in the orbital plane. The dispersion in the
third panel is on average constant: it is in the direction 
perpendicular to the plane of 
motion. Its periodic behaviour is due to the fact that we did not 
start with a diagonal matrix in action-angles.
The initial transformation from $({\bf x},{\bf v})$ to $({\bf \phi}, {\bf J})$
produces cross terms between all three directions. As the system 
evolves, and we project again onto configuration space, 
our 6-D ellipsoid rotates continually, 
producing a contribution in the 
direction perpendicular to the
orbital plane which varies with the frequencies 
$\Omega_r$ and $\Omega_\theta$.
By fitting $\sigma(v)/\sigma_0(v) = a/(1 + t/t_0)$, we find
for the velocity dispersion in the first panel $ a = 1.5$ and $t_0 = 0.6$ Gyr,
whereas for the dispersion in the second panel $ a = 2.6$ and $t_0 = 0.1$ Gyr.
\begin{figure*}
\label{f:vel}
\flushleft{\psfig{figure=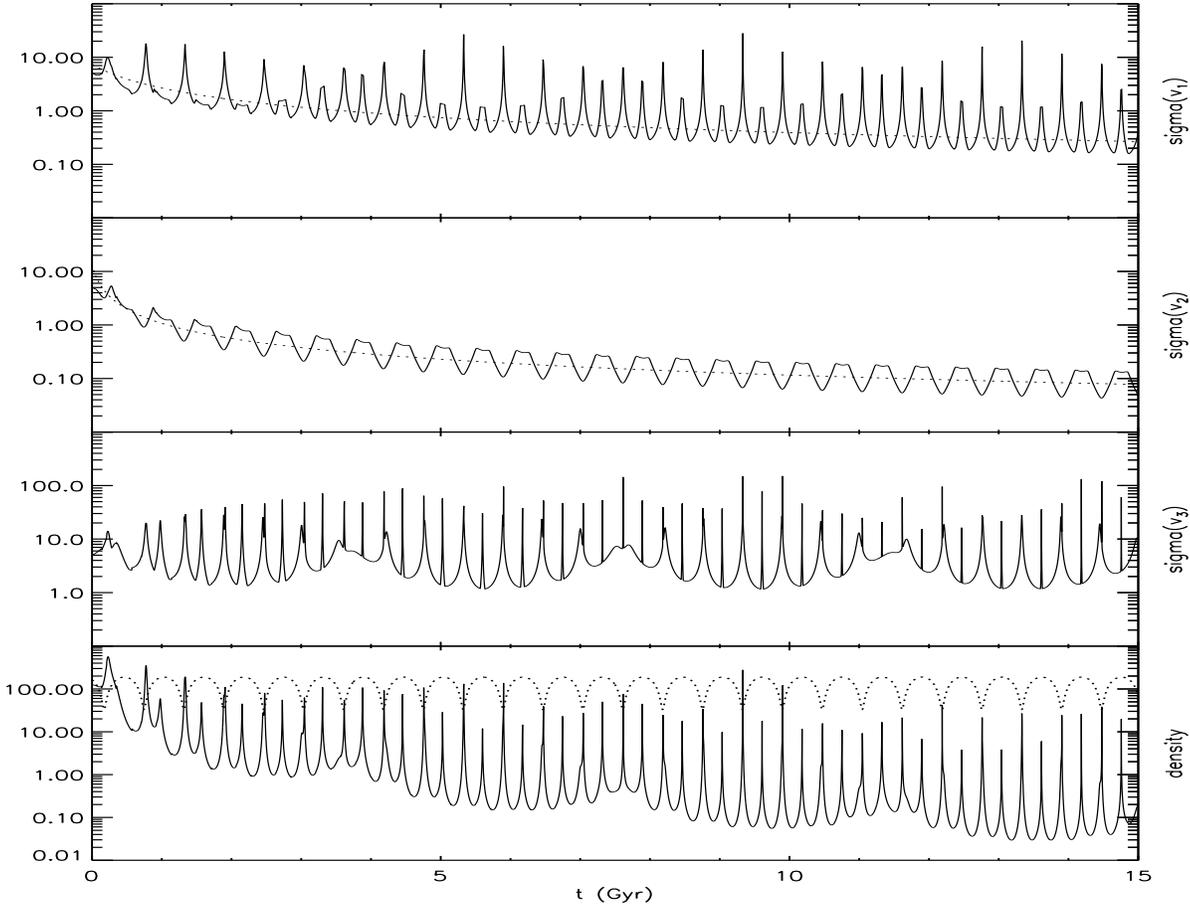,height=12.0cm,width=15.8cm}}
\caption[]{Time evolution  of the 
velocity dispersions along the major axis, computed as outlined in
Section~4.2, for the logarithmic spherical potential of Eq.~(\ref{eq:halo}). 
Two of the dispersions
decrease with time as $1/(1 + t/t_0)$ (dotted curve), 
whereas the third one is constant on the average. 
The periodic variations are due to the combination of
the radial and angular oscillations, as described in the text. 
The last panel shows the product of the three dispersions which is
proportional to the density (full curve). The radial oscillation is shown
(dotted curve) so that the occurrence of density spikes can be 
compared with the location of the turning points of the orbit.}
\end{figure*}

In the last panel we show the behaviour of
the product of the three dispersions, which is proportional to
the density (see Eq.~(\ref{eq:rho_sigma})). Note that, since two of the  
velocity dispersions have
decreased approximately a factor of ten, the density has done so
by a factor of hundred. Note also the decrease in the amplitude of
the spikes and the good correlation of these with the turning points
of the orbit. 

\subsubsection{Comparison to the simulations}

In order to assess the limitations of our approach, we will compare
our predictions with simulations of satellites with and without
self-gravity. We first consider what happens to a satellite with no
self-gravity moving in a spherical logarithmic potential. We take two
different sets of initial properties for the satellites: $1 \,{\rm kpc}$ 
width and
$\sigma_{1D} = 5 \,{\rm km}{\rm s^{-1}}$, corresponding to an initial
mass of $\sim 5.9 \times 10^7 \,{\rm M}_{\odot}$; and $5 \,{\rm kpc}$ width and
$\sigma_{1D} = 20\, {\rm km}{\rm s^{-1}}$, corresponding to  
$M \sim 4.7 \times 10^9 \,{\rm M}_{\odot}$ for the larger satellite.
Both begin as  
spherically symmetric Gaussians in coordinate and velocity space. 
We launch them on the same orbit so that we can directly study 
the effects of the change in size.

What observers measure are not the velocity dispersions or densities
of a stream at a particular point, but mean values given by a set of stars in a
finite region. We can estimate the effects of this
smoothing by comparing our analytic predictions with the simulations. In the
upper panel of Figure~8 we show the time evolution of the density 
(normalized to its initial value) 
for the small satellite. The full line represents 
our prediction and the stars correspond to the simulation.
We simply follow the central particle of the system as a function of time,
and count the number of particles 
contained in a cube of 1 kpc on a side surrounding it. 
Triangles represent the number density from an 8 times larger
volume (2 kpc on a side). 
The agreement between the
predictions and the estimated values from the simulations is very
good. The representation of a continuous field with a finite number of
particles introduces some noise which, together with the smoothing, is 
responsible for the disagreement. Note, however, how well the simulated
density spikes agree with those predicted at the orbital turning points. 
The overall agreement is slightly better for the small
cube than for the large one. This is due to the smoothing which inflates
some of the dispersions as a result of velocity gradients along
the stream.

In the lower panel of Figure~8 we show a similar comparison for the
large satellite. 
In general the prediction does very well here also. 
Note for the small boxes and at late times, we only have 
simulation points at the spikes (i.e. when the density is
strongly enhanced). 
This is because the satellite initially  has
a larger velocity dispersion and therefore spreads out more rapidly
along its orbit.
\begin{figure*}
\label{f:dens}
\flushleft{\psfig{figure=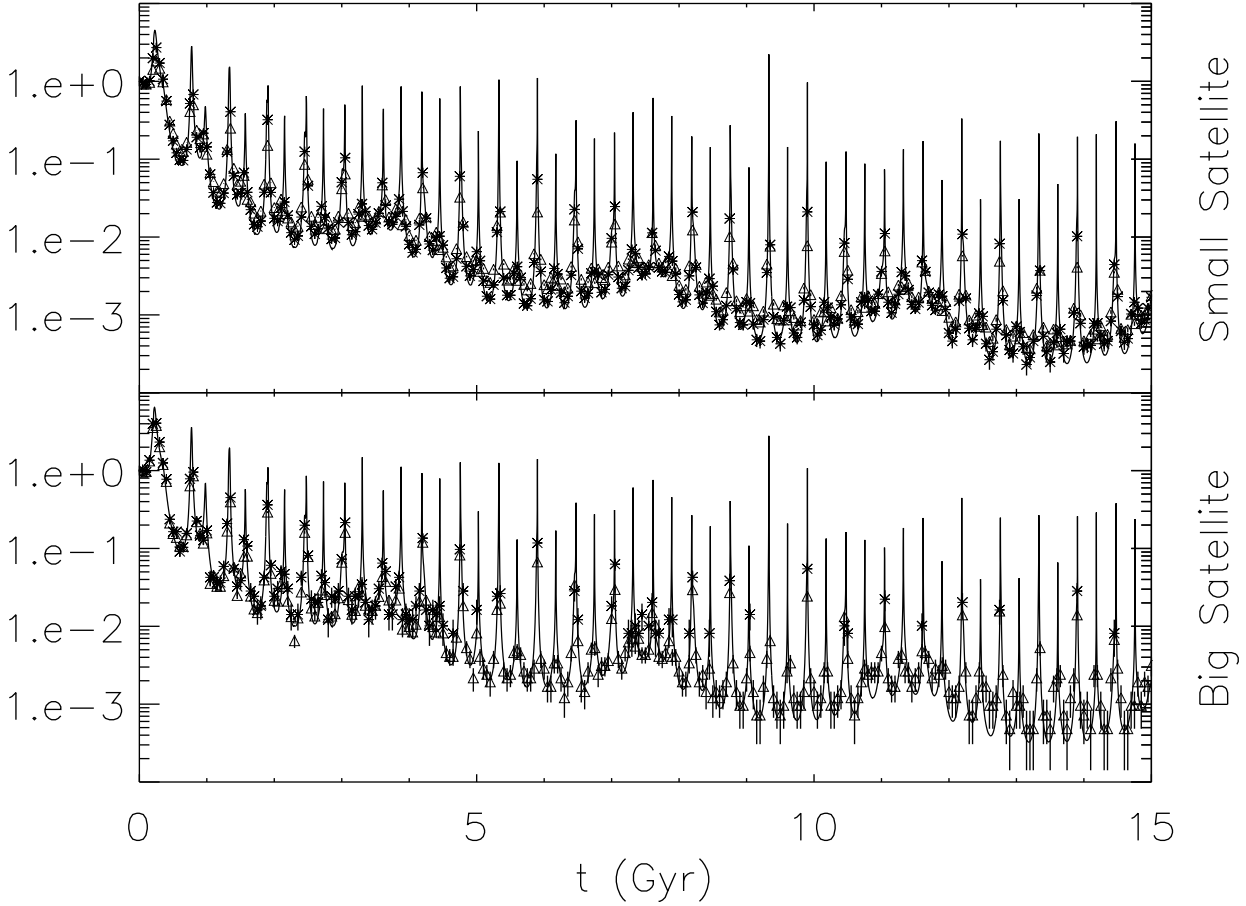,angle=90,height=12.0cm,width=16.4cm}}
\caption[]{Time evolution  of the density
for a satellite moving in a spherical potential (Eq.~(\ref{eq:halo})), 
with similar
orbital parameters as those of Experiment~6 in Table~1. 
The full line represents our prediction, normalized to the initial
density. In the upper panel we plot the density behaviour for the 
$\sim 5.9 \times 10^7 \sm$ satellite (see main text), 
whereas the lower panel corresponds to the $\sim 4.7 \times 10^9 \sm$ 
satellite. The stars indicate the number of particles that fall in a
volume of 1 kpc on a side around the central particle of the system, and the
triangles represent the number of particles in a cubic
volume of twice the side, both normalized to the initial value.
The spike-like behaviour occurs at the turning points of the orbit
(see main text -- Eq.~(\ref{eq:rho_sph})).}
\end{figure*}

We tested the effect of including self-gravity in the small
satellite simulation, and found no significant qualitative or quantitative
difference in the behaviour.

\subsection{Axisymmetric case}

As an illustrative example of the main characteristics of the
axisymmetric problem, let us consider the class of Eddington 
potentials $\Phi(r,\theta) = \Phi_1(r) + \eta(\beta \cos{\theta})/r^2$
(Lynden-Bell 1962; 1994) which are separable in spherical coordinates.
The third integral for this type of potentials is
$I_3 = \frac{1}{2} L^2 + \eta(\beta \cos{\theta})$. The actions are
computed from:
\begin{eqnarray}
J_\varphi \!\!\!& = &\!\!\! L_z, \\
J_\theta \!\!\!& = &\!\!\! \frac{1}{2 \pi}\oint   d\theta \,
\sqrt{2 (I_3 - \eta(\theta)) - \frac{J_\varphi^2}
{\sin^2{\theta}}}, \\
J_r \!\!\!& = &\!\!\! \frac{1}{2 \pi} \oint dr \,
\sqrt{2 (E - \Phi_1(r)) - \frac{2 I_3}{r^2}}.
\end{eqnarray}

Since the frequencies of motion are all different and non-zero, the system 
has the freedom to spread along three directions in phase-space. The
conservation of the local phase-space density will force the dispersions
along the remaining three directions to decrease in time.

Following a similar analysis as for the spherical case
we derive 
for the density at the central point ${\bf \bar{x}}(t)$ of the system at time
$t$ 
\begin{equation}
\label{eq:rho_ax}
\rho({\bf{\bar x}},t) =  \frac{(2 \pi)^{3/2} f_0}
{\sqrt{\det{\bf \sigma}_{\bf \phi}^{0}}}
\frac{1}{|\det{\bf \Omega'}|} 
\frac{\partial I_3}{\partial J_\theta} \frac{\Omega_r}
{r^2 \sin\theta |p_r p_\theta|}\frac{1}{t^3},
\end{equation}
where ${\bf \sigma}_{\bf \phi}^{0}$ is the angle submatrix of the 
initial variance matrix in action-angle variables. 
Therefore the density at the central point of the system
decreases as $t^{-3}$, because of the extra degree of freedom that
the rupture of the spherical symmetry introduces (see Appendix B), and
so after a Hubble time, the 
density decreases by approximately a factor of a thousand.

In Figure~9 we plot the time evolution of the components of
the velocity ellipsoid  for a
system on an orbit with the same
initial conditions as for the spherical case, in the potential 
\begin{equation}
\label{eq:axis_pot}
\Phi(r, \theta) = v_{\rm h}^2 \log{(r^2 + d^2)} + 
\frac{\beta^2 \cos^2\theta}{r^2}, 
\end{equation}
where $v_{\rm h} = 123$ \kms, $d = 12$ kpc and $\beta = 950$ kpc \kms.
This choice of parameters produces a reasonably flat potential 
which is physical (giving a positive density field) outside 
\mbox{7 kpc}. All velocity dispersions now decrease as $1/t$. 
\begin{figure*}
\label{f:velax}
\flushleft{\psfig{figure=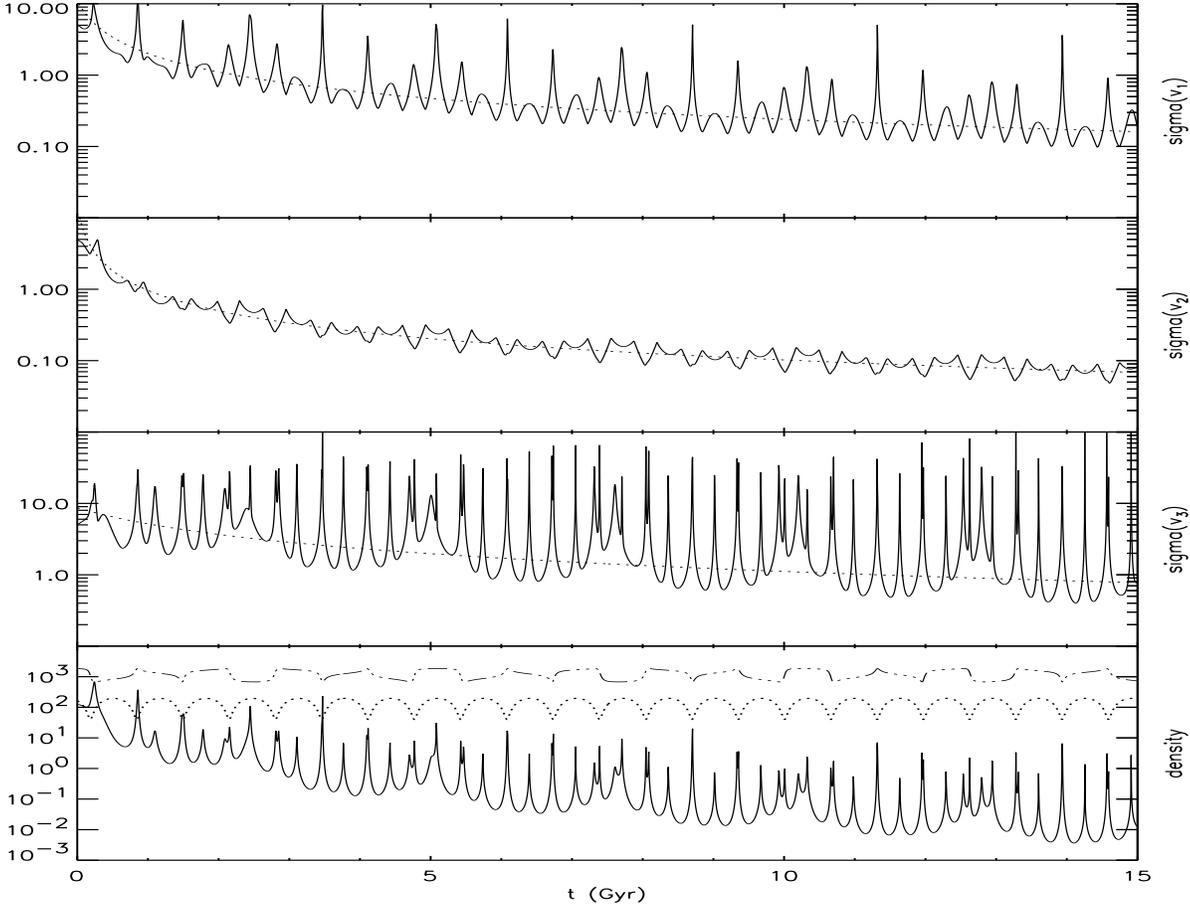,height=12.0cm,width=15.8cm}}
\caption[]{Time evolution  of the 
velocity dispersions along the principal axes, computed as outlined in
Section~4.2 and 4.3, for the simple axisymmetric potential of 
Eq.~(\ref{eq:axis_pot}). 
Now all the dispersions
decrease with time as $1/t$ (dotted curve).
The periodic time behaviour is due to the combination of
the radial and angular oscillations, as described in the text. 
The last panel shows the product of the three dispersions which is
proportional to the density. The radial and $\theta$-oscillations are
also plotted to indicate the position of the turning points.}
\end{figure*}

The analytic formalism developed here can be applied to any 
separable potential in a straightforward manner, 
using the definitions and results of
Sec.~\ref{sec:general}. This includes, of course, the set 
of St\"ackel potentials
which may be useful in representing the Milky Way (Batsleer \& Dejonghe 1994), 
or any axisymmetric
elliptical galaxy (de Zeeuw 1985, Dejonghe \& de Zeeuw 1987). 
The only difference is that 
the matrix ${\bf T}$ of the  transformation  from the usual 
coordinates (${\bf x}$,${\bf v}$) to the
action-angle variables 
should be first multiplied by the matrix of the mapping
from (${\bf x}$,${\bf v}$) to the ellipsoidal coordinates
$(\lambda,\mu,\varphi, p_\lambda,p_\mu,p_\varphi)$, since this is the
system in which the problem is separable. We discuss some of the properties
St\"ackel potentials and derive, for a particular model for our Galaxy,
the explicit form for the density in Appendix C.
Even if the potential is not
separable our general results on the evolution of the system remain
valid provided most orbits remain regular. In the general case  
the frequencies and their derivatives with
respect to the actions will have to be computed through
a spectral dynamics analysis similar to that used in 
Section 3.1 (Carpintero and Aguilar 1998).

\subsection{What happens if there is phase-mixing}

The procedure outlined above assumes that only one stream
of debris from the satellite is present in any volume which is
analysed. When phase-mixing becomes important 
we may find more than one kinematically cold stream near a given 
point. The velocity dispersions of the debris in such a region
would then appear much larger than predicted naively
using our formalism. 
We can make a rough estimate for the velocity dispersions also 
in this case by using the following simple argument.

If the system is (close to) completely phase-mixed, then
the coarse-grained distribution function that describes it 
will be uniform in the angles and therefore will
only depend on the adiabatic invariants, i.e. $f({\bf x}, {\bf v}) =
f({\bf J}({\bf x}, {\bf v}))$. Since these are conserved 
the moments of the coarse-grained distribution function will be 
given by the moments of the initial distribution function. 
Therefore $f(\bf{J})$ is completely determined by the initial properties of
the system in the adiabatic invariants space. If the initial distribution
function is Gaussian in action-angles then $f(\bf{J})$ will be Gaussian
with mean and dispersion given by their values at $t=t^0$. 

As an example, let us analyse the velocity dispersion in the 
$\varphi$-direction in a 
particular region in which there is a multistream structure:
\[
\sigma^2(v_\varphi) = \frac{\int d^3x~ d^3 v 
~(v_\varphi - \bar{v}_\varphi)^2 
~f(\bf{J}(\bf{x}, \bf{v}))}{\int d^3x ~d^3 v~ f(\bf{J}(\bf{x}, \bf{v}))} 
 = \frac{\D\int d^3x~ d^3J~ \D\left(\frac{{J}_\varphi}{R} - 
\frac{\bar{J}_\varphi}{\bar{R}}\right)^2 f(\bf{J})}{\int d^3x~ d^3J~f(\bf{J})},
\]
where we used that $v_\varphi = {J}_\varphi/R$. By expanding to first order
we find
\begin{equation}
\label{eq:sLz}
\sigma^2(v_\phi) = \sigma^2(J_\varphi)/\bar{R}^2 + 
\Delta_x^2 \bar{J}_\varphi^2/\bar{R}^4.
\end{equation}\noindent
Here we replaced $\sigma(R)$ by $\Delta_x$ (the size of 
the region in question) which is justified by our previous result that
the spatial dimensions of streams grow with time; 
and neglected the correlation between 
$J_\varphi$ and $R$. The first term in Eq.~(\ref{eq:sLz}) estimates 
the dispersion between streams, while
the second estimates the contribution from the velocity gradient
along an individual stream.
For the experiments of Table~1 the values of the dispersions 
range from 50 to 150 \kms. These dispersions increase in proportion 
to those of the initial satellite. 

\subsubsection{The filling factor}

We can use the results of our previous section to 
quantify the probability of finding 
more than one stream at a given position in space. 
This probability is measured by the filling factor. We define this 
by comparing the mass-weighted spatial density
of individual streams 
with a mean density estimated by dividing the mass of the
satellite by the total volume occupied by its orbit. The first
density can be calculated formally through an integral over the
initial satellite:
\[
 \langle \, \rho(t)\, \rangle = \frac{1}{M} 
\int dm({\bf x,v}) ~\rho({\bf x,v})(t) 
	= \frac{1}{M} \int 
d^3x~d^3v f({\bf x,v}, t^0)~\rho({\bf x,v})(t),\nonumber
\]
where $\rho({\bf x,v})(t)$ is the density at time $t$ of the
individual stream in the neighbourhood of the particle which
was initially at $({\bf x,v})$. The filling factor is then
\[F(t) = \frac{M}{V_{o}}\frac{1}{\langle \, \rho(t)\, \rangle},\]
where $V_{o}$ is the volume filled by the satellite's orbit.
An estimate of the filling factor can be obtained by
approximating $\langle \, \rho(t)\, \rangle$ by 
$\rho(\bar{\bf x},t)/(2 \sqrt{2})$ taken  
from Eqs.~(\ref{eq:rho_sph}), (\ref{eq:rho_ax}) or (\ref{eq:rho_staeckel}) 
for spherical, axisymmetric Eddington or St\"ackel potentials respectively.
The factor $1/2 \sqrt{2}$ is the ratio of the central to mass-weighted 
mean density for a Gaussian satellite. 
We approximate $V_o = 4\pi\, r_{\rm apo}^3 \cos \theta_{\rm f}/3 $,
where $r_{\rm apo}$ and $\theta_{\rm f}$ correspond to  the orbit of the 
satellite centre. Since we are interested in deriving an estimate for
the filling factor for the solar neighbourhood, we focus 
on the St\"ackel potential described in Appendix C, which produces
a flat rotation curve resembling that of the Milky Way. Thus
\begin{equation}
F(t) =\frac{6 \sqrt{2} M \sqrt{\det{\bf \sigma_\phi^0}}}
{2 (2 \pi)^{5/2} \, f_0} 
\frac{\langle \, R\, \rangle  \langle \, |\nu - \lambda|v_\lambda v_\nu\, 
\rangle }{r_{\rm apo}^3 \cos{\theta_{\rm f}}}  \frac{|\det{\bf \Omega'}|}
{\D\left|\Omega_\nu \frac{\partial I_3}{\partial J_\lambda} - 
\Omega_\lambda \frac{\partial I_3}{\partial J_\nu}\right|} t^3, 
\end{equation}
where $\lambda$, $\nu$ are spheroidal coordinates (for which the potential is
separable), $J_\lambda$ and $J_\nu$ are the corresponding actions, and
$\Omega_\lambda$ and $\Omega_\nu$ the frequencies; and $I_3$ is the third 
integral of motion. If we approximate $\langle \, v_\lambda v_\nu\, 
\rangle  \sim v_{\rm circ}^2/4$
and replace $f_0 = M/(2 \pi \sigma(x) \sigma(v))^3$ then
\begin{equation}
\label{eq:fil_fac_gral}
F(t) \sim C_{\rm orbit} C_{\rm IC}
 \left(\frac{\sigma(x)}{r_{\rm apo}}\right)^2 \,
\frac{\sigma(v)}{v_{\rm circ}} \,\left(\Omega_\lambda \,t\right)^3,
\end{equation}
where
\begin{equation}
 C_{\rm orbit} = \frac{3 \sqrt{\pi} ~\langle \, |\nu - \lambda|\, \rangle ~
\langle \, R\, \rangle ~v_{\rm circ}^5 ~|\det{\bf \Omega'}|}
{ 2 \cos{\theta_{\rm f}}\D\left|\Omega_\nu \frac{\partial I_3}
{\partial J_\lambda} - 
\Omega_\lambda \frac{\partial I_3}{\partial J_\nu}\right|
\Omega_\lambda^3},
\end{equation}
depends on the orbital parameters of the satellite, and
\begin{equation}
C_{\rm IC} = \frac{h_\lambda 
h_\nu}{\D\left|\Omega_\nu \frac{\partial I_3}{\partial J_\lambda} - 
\Omega_\lambda \frac{\partial I_3}{\partial J_\nu}\right|} 
\frac{\lambda-\nu}{P^3 Q^3}
\frac{R}{r_{\rm apo} v_{\rm circ}^2}\Bigg\rfloor_{{\bf \bar{x}^0}, 
{\bf \bar{v}^0}},
\end{equation}
with
\[ h_\tau = 2 p_\tau \frac{\partial p_\tau}{\partial \tau}, 
\qquad \tau = \lambda, \nu,\]
is a function of its initial position on the orbit. 
(See Appendix C for further details and definitions). 
This last expression holds if
the satellite is initially close to a turning point of its orbit. 

For example, a satellite of 10 \kms velocity dispersion and
$0.4$ kpc size on an orbit with an 
apocentric distance of 13 kpc, a maximum  height above the plane of 
5 kpc and an orbital period of $\sim$ 0.2 Gyr,
gives an average of $0.4$ streams of stars 
at each point in the inner halo after 10 Gyr. 
A satellite of 25 \kms dispersion and 1 kpc size  on the
same orbit would produce $5.9$ streams on the average after the same time. 
Let us compare this last prediction with a simulation for the same satellite
and the same initial conditions
in the Galactic potential described in Section~2. 
In Figure~10 we plot the behaviour of the filling factor
from the simulation, computed as 
\[ F(t) = \frac{N}{V_o} \frac{1}{n(t)}, \]
where $N$ is the total number of particles, 
$n(t) = N^{-1} \sum_i \rho_i$
with $\rho_i$  the density of the stream where particle $i$ is, which we
calculate by dividing space up into
2 kpc boxes and counting the number of particles of each
stream in each box. Note that the filling factor increases as $t^3$
at late times as we expect for any axisymmetric potential. 
Our prediction is in good
agreement with the simulations, showing also that it is 
robust against small changes in the form of  
the Galactic potential.
\begin{figure}
\center{\psfig{figure=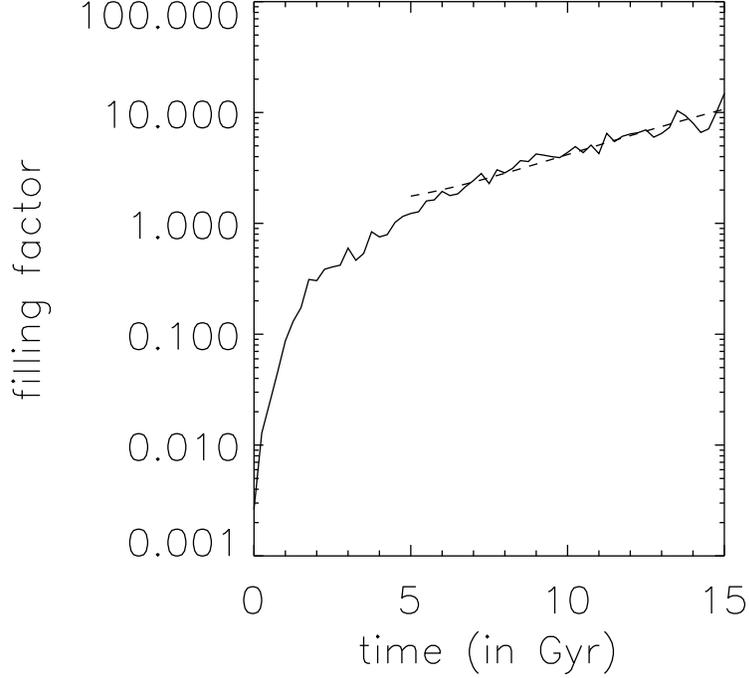,height=9.0cm}}
\caption[]{Time evolution  of the filling factor for a satellite
with an initial velocity dispersion of 25 \kms and size of 1 kpc, 
moving in the Galactic potential described in Section~2. Its orbital
parameters resemble those of halo stars in the solar neighbourhood. 
The dashed-curve indicates a $\gamma_0 + \gamma_1 t^3$ fit
for late times.}
\end{figure}

\subsubsection{Properties of an accreted halo in the solar neighbourhood}

To compare with the stellar halo it is more useful 
to derive the dependence of the filling
factor on the initial luminosity of a satellite. We shall assume that
the progenitor satellites are similar to present-day dwarf ellipticals, 
and satisfy both a Faber-Jackson relation:
\begin{equation}
\label{eq:faber_jackson}
 \log \frac{L}{\rm L_\odot} - 3.53 \log \frac{\sigma(v)}{\rm km \, s^{-1}}
 \sim 2.35, 
\end{equation}
for $H_0 \sim 50 \,{\rm km \,s^{-1} Mpc^{-1}}$, and a 
scaling relation between the effective radius 
($R_e \sim \sigma(x)$) and the velocity dispersion $\sigma(v)$:
\begin{equation}
\label{eq:effrad_sigmavel}
 \log \frac{\sigma(v)}{\rm km \, s^{-1}} - 1.15 \log \frac{R_e}{\rm kpc} 
\sim 1.64, 
\end{equation}
both as given by  Guzm\'an, Lucey \& Bower (1993) for
the Coma cluster.  
Expressed in terms of the luminosity of the progenitor, the filling factor
then becomes
\begin{equation}
\label{eq:filfac_L}
F(t) \sim C_{\rm orbit}\, C_{\rm IC} \left(\frac{L}{L_n}\right)^{0.776}\,
(\Omega_\lambda \, t)^3, 
\end{equation}
where $L_n$ is a normalization
constant that depends on the orbit and
on the properties of the parent galaxy as:
\begin{equation}
L_n = 3.75 \times 10^{11} {\rm L_\odot} \, 
\left(\frac{r_{\rm apo}}{10~ {\rm kpc}}\right)^{2.58}
\left(\frac{v_{\rm circ}}{200~ {\rm km\, s^{-1}}}\right)^{1.29}.
\end{equation}

If the whole stellar halo had been built from disrupted satellites, 
we can derive the number of streams expected in 
the solar neighbourhood by adding their filling factors 
using the appropriate orbital parameters in Eq.~(\ref{eq:fil_fac_gral}) or
Eq.~(\ref{eq:filfac_L}): $F_\odot(t) = N_{\rm sat} F(t)$.
 For a sample of giant stars located within 
1 kpc from the Sun with photometric distances and 
radial velocities measured from the ground 
(Carney \& Latham 1986; Beers \& Sommer-Larsen 1995; Chiba \& 
Yoshii 1998), and proper motions measured by HIPPARCOS,  we estimate  
$C_{\rm orbit} \times C_{\rm IC} \sim 1.29 \times 10^{-3}$. The
median pericentric (apocentric) distance is $3.7$ ($ 11.6$) kpc,
and the median $\Omega_\lambda$ is 26.6 Gyr$^{-1}$ (equivalent to a period
of $\sim 0.24$ Gyr). Thus using Eq.~(\ref{eq:fil_fac_gral}) 
\[
 F_\odot(t) \sim 0.9 N_{\rm sat} \left(\frac{\sigma(x)}{\rm kpc}\right)^2 \,
\frac{\sigma(v)}{\rm km~s^{-1}} \,
\left(\frac{t}{10 \,{\rm Gyr}}\right)^3.
\]
If now we assume that the progenitor systems are similar to present-day dwarf
ellipticals, then using Eq.~(\ref{eq:filfac_L}) we find
for the whole $10^9~{\rm L}_\odot$ stellar halo 
\begin{equation}
 F_\odot(t) \sim \left(\frac{t}{10 \,{\rm Gyr}}\right)^3 \times 
\left\{\begin{array}{lr} 5.1 \times 10^{2}, &
 \mbox{100}\times  10^7 ~{\rm L}_\odot \,\, {\rm  sat}, \\
 3.0 \times 10^{2}, &
 \mbox{ 10} \times 10^8 ~{\rm L}_\odot \,\,{\rm  sat}.
\end{array}\right.
\end{equation}
For $t \sim 10$ Gyr, the {\em number of streams}
expected in the solar neighbourhood is therefore in the range
\begin{equation}
F_{\odot} \sim 300 - 500.
\end{equation}

Fuchs \& Jahrei\ss\ (1998) have obtained a lower limit for 
the local mass density of spheroid
dwarfs of $1 \times 10^{5} \sm {\rm kpc}^{-3}$.
We may use this estimate to derive the mass content in subdwarfs of 
an individual stream in a volume of 1 kpc$^3$ centered on the Sun:
\begin{equation}
F_{M}(t) \sim \frac{M_{\rm local~halo} ~({\rm in~ 1~ kpc}^3)}{F(t)}.
\end{equation}
Thus with our previous estimate for the filling factor 
\begin{equation}
F_{M}(t) \sim \left(\frac{10 ~{\rm Gyr}}{t}\right)^3 \times
\left\{\begin{array}{ll} 1.9 \times 10^2 \,{\rm M}_\odot, &
{\rm for}\, 10^7 ~{\rm L}_\odot \,{\rm sat,} \\
3.3 \times 10^2 \,{\rm M}_\odot, &
 {\rm for}\, 10^8 ~{\rm L}_\odot \,{\rm sat}.
\end{array}\right.
\end{equation}
Therefore, after 10 Gyr, each stream contains  
$F_M \sim (200 - 350) \,{\rm M}_\odot$ in subdwarf stars, 
depending on the orbital parameters of the progenitors and their
initial masses.  

Since the halo stars in the solar neighbourhood
have one-dimensional dispersions 
$\sigma_{\rm obs}(v) \sim 100 - 150$ \kms, in order to 
distinguish kinematically whether their distribution 
is really the superposition of $\sim 300 - 500$ individual 
streams of velocity dispersion $\sigma_{\rm st}(v)$ we might
require that 
\begin{equation}
\sigma_{\rm st}^{3}(v) <  \frac{1}{27}\frac{\sigma_{\rm obs}^3(v)}{F_{\odot}},
\end{equation}
where the factor $1/27$ would ensure a $\sim 3 \sigma$ distinction between
streams. Using our previous estimate of $F_{\odot}$ this condition 
becomes \mbox{$\sigma_{\rm st}(v) < 
\sigma_{\rm obs}(v)/(20 - 24)$}, and thus 
\mbox{$\sigma_{\rm st}(v) < 5$ \kms}.
Currently the observational errors in the measured velocities 
of halo stars are of order 20 \kms, and thus 
there is little hope to distinguish at the present day 
all the individual streams which may make up 
the stellar halo of our Galaxy. 
Since intrinsic velocity dispersions for streams originating from
$10^7 - 10^8 {\rm L}_\odot$ objects are of the order 
of $3 - 5$ \kms after 10 Gyr, 
it should be possible to distinguish such streams
with the astrometric missions SIM and GAIA, if they reach their
planned accuracy of a few \kms. Even with an accuracy of 15 \kms 
per velocity component, streams are predicted to be marginally separated.
The clumpy nature of the distribution should thus be easily distinguishable in
samples of a few thousand stars. 
One way of identifying streams which are debris from the same 
original object, is through clustering 
in action or integrals of motion space (Helmi, Zhao \& de Zeeuw 1998). 

\section{An observational application}
  
Majewski et al. (1994) discovered a clump of nine halo stars 
in a proper motion survey near the NGP (Majewski 1992), which 
appeared separated from the main distribution of stars in the field. They
measured proper motions, photometric parallaxes, $F$ magnitudes and 
$(J-F)$ colours 
for all nine stars and radial velocities for six of them. 
For these six stars we find for the mean velocity 
$\bar{v}_\varphi = -152 \pm 23 
\, {\rm km\,s^{-1}}$, 
$\bar{v}_R = -260 \pm 18 \, {\rm km\,s^{-1}}$ and 
$\bar{v}_z = -76 \pm 18\, {\rm km\,s^{-1}}$, and for 
the velocity dispersions 
$\sigma(v_{\varphi}) =  99 \pm 33\, {\rm km\,s^{-1}}$, 
$\sigma(v_R) = 100 \pm 24 \,{\rm km\,s^{-1}}$ and
$\sigma(v_z) = 35 \pm 24 \,{\rm km\,s^{-1}}$. If the 
dispersions are computed  along the principal axes, we find  
$ \sigma(v_1) = 29 \pm 20 \,{\rm km\,s^{-1}}$, $
\sigma(v_2) = 68 \pm 94 \,{\rm km\,s^{-1}}$, 
$\sigma(v_3) = 125 \pm 5 \,{\rm km\,s^{-1}}$.

Since the mean velocities are significantly different from zero, the group
of stars can not be close to any turning point of their orbit. 
The only way to understand the large observed
dispersions, in particular of $\sigma(v_3)$, if the stars come from a
single disrupted satellite, 
is for the group to consist of more than one stream of stars. 
We believe that this may actually be the case. 
By computing the angular momenta
of the stars we find they
cluster into two clearly distinguishable subgroups:
$\bar{L}_z^{(1)} =  -784$ and $\sigma^{(1)} (L_z) = 299$, and 
$\bar{L}_z^{(2)} =  -2180$ and $\sigma^{(2)} (L_z) = 313$ in kpc \kms.
If we accept the existence of two streams as
a premise, we may compute the velocity dispersions in each of them. 
We find for the stream with 4 stars
\[ \sigma^{(1)}(v_1) = 25 \pm 25 , \, \,
\sigma^{(1)}(v_2) = 43 \pm 62 , \, \, 
\sigma^{(1)}(v_3)= 100 \pm 45,
\]
while for the stream with 2 stars
\[ \sigma^{(2)}(v_1) = 3 \pm 4 , \, \,   
\sigma^{(2)}(v_2) = 25 \pm 21 , \, \,  
\sigma^{(2)}(v_3) = 89 \pm 64,
\]
all in \kms. 
These results are consistent at a $2\sigma$ level 
with very small 3-D velocity dispersions,  
as expected, if indeed these are streams 
from a disrupted satellite.  

With this interpretation  of the kinematics of this group, we 
can estimate the mass of the progenitor and its initial
size and velocity dispersion. 
Galaxies today obey scaling laws of the Faber-Jackson or
Tully-Fisher type. If we assume that the original satellite was similar to 
present-day dwarf ellipticals, 
then we may use Eq.~(\ref{eq:effrad_sigmavel})
to derive a relation between the initial dispersion in the 
$z$-component of the angular momentum and initial velocity dispersion
of the progenitor
\begin{equation}
\sigma_i^2(L_z) = \sigma_i^2(v) R_{\rm apo}^2 + 
0.0375^2 \frac{L_z^2}{R_{\rm apo}^2} \sigma_i^{1.74}(v), 
\end{equation}
where $R_{\rm apo}$ is the apocentric distance of its orbit.  
Under the assumption that
$L_z$ is conserved, we can derive $\sigma_i(v)$ by replacing
in the previous equation the observed values of $L_z$, $\sigma(L_z)$
and an estimate of $R_{\rm apo}$. We obtain the latter
by orbit integration in a Galaxy model, which includes
a disk, bulge and halo and find $R_{\rm apo} \sim 12 \,{\rm kpc}$. 
Our estimate for the initial velocity dispersion of the progenitor is then
\begin{equation}
\sigma_i(v) \sim 48 \,{\rm km \,s}^{-1},
\end{equation}
which in Eqs.~(\ref{eq:faber_jackson}) and (\ref{eq:effrad_sigmavel})
yields for its initial luminosity and size
\begin{equation}
L \sim 2 \times 10^8 ~{\rm L}_{\odot}, \qquad R \sim 1 \,{\rm kpc}. 
\end{equation}
We estimate that the relative error-bars in these quantities are
 of order 50\%, 
if measurement errors and a 50\% uncertainty in the apocentric distance
are included. 

In summary, if indeed these stars come from a single disrupted
object, we must accept that the first six stars that were detected
(Majewski et al. 1994) 
are part of at least two independent streams. This seems reasonable, 
since two streams can be indeed be distinguished, and 
the velocity dispersions, in each stream are very small. 
Moreover, a disrupted object with the properties just derived 
(luminosity, initial size
and velocity dispersion), would fill its available volume rapidly, 
producing a large number of streams. 
In view of our explanation, a number of stars
from the same disrupted object but 
with positive $z$-velocities should also be present in the same region,
since phase-mixing allows streams to be observed with
opposite motion in the $R$ and/or $z$ directions. Candidates
for such additional debris should have similar
$v_\varphi$, since $L_z$ is conserved during 
phase-mixing. By simple inspection 
of Figure~1(a) in Majewski et al. (1994), 
other stars can be indeed found, with similar $v_\varphi$ but 
opposite $v_R$ and $v_z$.

\section{Discussion and Conclusions}

We have studied the disruption of satellite
galaxies in a disk + halo potential and 
characterised the signatures 
left by such events in a galaxy like our own. 
We developed an analytic description 
based on Liouville's theorem and on the 
very simple evolution of the system in action-angle
variables. This  is applicable to any accretion
event if self-gravity is not
very important and as long as the overall potential is static or 
only adiabatically changing.
Satellites with masses up to several times  $10^9 \, {\rm M}_{\odot}$
are likely to satisfy this adiabatic  condition if the mass of the Galaxy
is larger than several times $10^{10} \, {\rm M}_{\odot}$  at
the time of infall and if there are no other strong perturbations.
Even though have not studied how the system gets to its starting point, 
it seems quite plausible that in this regime dynamical friction will
bring the satellites to the inner regions of the 
Galaxy in a few Gyr, where they will be disrupted very rapidly. 
Their orbital properties may be similar to those found in 
CDM simulations of the infall structure onto clusters, where objects are 
mostly on fairly radial orbits (Tormen, Diaferio \& Syer 1998);
this is consistent with the dynamics of solar neighbourhood 
halo stars. Their masses 
range from the low values estimated observationally for dwarf spheroidals 
to the much larger values expected for the building blocks in
hierarchical theories of galaxy formation. 

We summarize our conclusions as follows. 
After 10 Gyr we find no strong 
correlations in the spatial distribution 
of a satellite's stars, since for orbits relevant to the bulk of the
stellar halo this is sufficient time for the stars to fill 
most of their available configuration 
volume. This is consistent with the fact that 
no stream-like density structures have so far been observed in the 
solar neighbourhood. 
On the contrary, strong correlations are present in velocity space.
The conservation of phase-space density 
results in velocity dispersions at each point along a stream
that decrease as $1/t$.  
On top of the secular behaviour, periodic oscillations are also expected: 
at the turning points of the orbit the velocity dispersions, and thus 
the mean density of the stream, can be considerably enhanced. 
Some applications of this density enhancement deserve
further study. For example, the present
properties of the Sagittarius dwarf galaxy 
seem difficult to explain, since numerical simulations 
show that it could have been disrupted very rapidly
given its current orbit (Johnston, Spergel \& Hernquist 1995;
Vel\'azquez \& White 1995). This puzzle has led to some unconventional
suggestions to explain its survival, like a massive and dense
dark matter halo (Ibata \& Lewis 1998) or a recent collision with the
Magellanic Clouds (Zhao 1998). However, since the densest part of
Sagittarius seems to be near its pericentre, it could be located sufficiently
close  to a `caustic' to be interpreted as a transient enhancement. 
Sagittarius could simply be a galaxy disrupted several
Gyr ago (c.f. Kroupa 1997). 

If the whole stellar halo of our Galaxy was built by merging of 
$N_{\rm sat}$ similar smaller systems of characteristic size $\sigma(x)$
and velocity dispersion $\sigma(v)$, then after 10 billion 
years we expect the stellar distribution in the solar 
neighbourhood to be made up of $F_\odot$ streams, where
\[ F_\odot \sim 0.9 N_{\rm sat} \left(\frac{\sigma(x)}{\rm kpc}\right)^2 \,
\frac{\sigma(v)}{\rm km~s^{-1}}.\]
For satellites which obey the same scaling relations as the dwarf elliptical
galaxies, this means 300 to 500 streams. Individually, 
these streams should have extremely small velocity dispersions,
and inside a 1 kpc$^3$ volume centered on the Sun each should 
contain a few hundred stars.  
Since the local halo velocity
ellipsoid has dispersions of the order of 100 \kms, 3-D  
velocities with errors smaller than 5 \kms are needed 
to separate unambiguously the individual streams. 
This is better by a factor of four 
than most current measurements, which 
would, however, be good enough to give a clear detection of the expected
clumpiness in samples of a few thousand stars.  
The combination of a strongly mixed population with relatively 
large velocity errors
yields an apparently smooth and  Gaussian
distribution in smaller samples. 
Since the intrinsic
dispersion for a stream from an LMC-type progenitor  
is of the order of $3 - 5$ \kms~ after a Hubble time, 
one should aim for velocity uncertainties below 3 \kms. 
With the next generation of astrometric satellites, (in particular GAIA, e.g.
Gilmore et al. 1998) 
we should be able to  distinguish almost all 
streams in the solar neighbourhood
originating from disrupted satellites.

Our analytic approach is based on 
Liouville's Theorem and  the very simple evolution 
of the system in action-angle
variables. Although the latter is likely to fail in the full merging
regime, the conservation of local phase-space density will
still hold. It will be interesting to see how this conservation law
influences the final phase-space distribution in the
merger of  more massive disk-like systems. These are plausible  
progenitors for the bulge of our Galaxy in hierarchical models. 

\section*{Acknowledgments}
A.H. wishes to thank the hospitality of MPA, HongSheng Zhao 
for many very useful discussions, Tim de Zeeuw for comments on earlier
versions of this manuscript and Daniel Carpintero for kindly providing
the software for the spectral-analysis used in Section~3.1.
EARA has provided financial support for A.H. visits to MPA.

\appendix 
\section{Spherical Potential}
\subsection{2-D case}

For a spherical potential $\Phi(r)$, the Hamiltonian is 
separable in spherical coordinates
and depends on the actions $J_\phi$ and $J_\theta$ only through the
combination $J_\phi + J_\theta = L$. We therefore
may choose a system of coordinates which coincides with the plane of
motion of the system, reducing the problem to 2-D. The position of a particle 
is given by its angular $\psi$ and radial $r$ coordinates on that plane. 
In that case, we have
\begin{equation}
\label{eq_apjr}
L = J_\psi = p_\psi, \qquad
J_r = \frac{1}{\pi} \int_{r_1}^{r_2} dr \frac{1}{r}
\sqrt{2 (E - \Phi(r)) 
r^2 - L^2}
\end{equation}
where $ L$ is the total angular momentum of the particle, $E$ its energy
and $r_1$ and $r_2$ are the turning points in the radial direction of motion. 
The action $J_r$ cannot be computed analytically in general 
for an arbitrary potential. 
However, Eq.~(\ref{eq_apjr}) defines an implicit function 
$g = g(E, L, J_r) \equiv 0$, 
which we can differentiate to find the frequencies of motion and their
derivatives. These are needed to
compute the elements of the matrix ${\bf \Theta}(t)$ 
(Eq.~(\ref{eq:matrix_th})) and to obtain the
time evolution of the distribution function. 

To simplify the computations, 
we assume that the variance matrix \footnote{As we mentioned in Section 4.2, 
${\bf \sigma}$ is in fact the inverse of the covariance
matrix. However we refer to ${\bf \sigma}$ as the 
variance matrix.} in action-angle variables
is diagonal at $t=0$: $ {\sigma^0}_{w_{ij}} = \sigma_{ii} {\bf \delta}_{ij}$. 
The evolution
of the system in phase space is obtained through the product 
${\bf \Theta}(t)^\dagger {\bf \sigma^0}_w {\bf \Theta}(t)$, 
which yields the following 
variance matrix $\sigma_w(t)_{ij} = \{i,j\}$ at time $t$
\[
{\bf \sigma}_w(t) = \left[ 
\begin{array}{cccc}
 \sigma_{11} & 0 &  -\sigma_{11}\Omega'_{11}t  &
 -\sigma_{11}\Omega'_{12}t \\
 \{1,2\} & \sigma_{22} &  -\sigma_{22}\Omega'_{12}t &    
 - \sigma_{22} \Omega'_{22} t \\
 \{1,3\} & \{2,3\} & \sigma_{11}{\Omega'_{11}}^{2} t^2 +  
	\sigma_{22}{\Omega'_{12}}^{2} t^2 + \sigma_{33} & 
 \Omega'_{11} \sigma_{11}\Omega'_{12}t^2 + 
\Omega'_{12} \sigma_{22}\Omega'_{22}t^2 \\
\{1,4\} & \{2,4\} & \{3,4\} &
\sigma_{11}{\Omega'_{12}}^{2}t^2 + \sigma_{22}{\Omega'_{22}}^{2}t^2 + 
 \sigma_{44} \end{array}\right]
\]
in action-angle variables, with $\Omega'_{ij} = 
\partial \Omega_i/\partial J_j$. Subindices $\{1\}$ and $\{3\}$ refer to
directions associated to $\psi$, such as for example $\phi_\psi$ and 
$J_\psi$, whereas $\{2\}$ and $\{4\}$ are related to $r$.

We find the properties of the debris in configuration and
momenta space by transforming the action-angle coordinates locally
around ${\bf \bar{x}}$ with the matrix ${\bf T}^{-1}$. Its elements are
the second derivatives of the characteristic
function $W(\bf{q},\bf{J})$:
\begin{equation}
\label{eq_aptransf}
{\bf T}^{-1} = \left[\begin{array}{cc} {\bf W}_{JJ} {\bf J}_q + {\bf W}_{Jq} 
& {\bf W}_{JJ} {\bf J}_p\\
{\bf J}_q & {\bf J}_p \end{array}\right]
\end{equation}
with ${\bf J}_q = - {\bf W}_{qJ}^{-1} {\bf W}_{qq}$ and 
${\bf J}_p =  {\bf W}_{qJ}^{-1}$,
and has the following form for a spherical
potential in 2-D
\begin{equation}
\label{eq_aptransf_el}
{\bf T}^{-1} =  \left[ 
{\begin{array}{cccc}
1 &  t_{12} &  t_{13} &  t_{14} \\
0 &  t_{22} &  t_{23} &  t_{24} \\
0 & 0 & 1 & 0 \\
0 &  t_{42} &  t_{43} &  t_{44}
\end{array}}
 \right] 
\end{equation}
with
\[\begin{array}{lll}
t_{12} = - {\D \frac{h(r)}{\Omega_r} 
\frac{\partial^2 W}{\partial L \partial J_r} + \frac{1}{p_r}\left(\Omega_\psi -
\frac{L}{r^2}\right),} &
t_{13} = {\D \frac{\partial^2 W}{\partial L^2} + 
\frac{\partial^2 W}{\partial L \partial J_r} t_{43}}, &
t_{14} ={\D\frac{\partial^2 W}{\partial L \partial J_r}\frac{p_r}{\Omega_r}}, 
\\
t_{22} = - {\D \frac{h(r)}{\Omega_r}  
\frac{\partial^2 W}{\partial J_r^2}
+  \frac{\Omega_r} {p_r}}, & 
t_{23} = {\D \frac{\partial^2 W}{\partial L \partial J_r} +
\frac{\partial^2 W}{\partial J_r^2} t_{43}}, &
t_{24} = {\D \frac{\partial^2 W}{\partial J_r^2}\frac{p_r}{\Omega_r}}, \\
t_{42} = {\D  - \frac{h(r)}{\Omega_r}}, & 
t_{43}  = {\D  - \frac{1}{\Omega_r}\left(\Omega_\psi -
\frac{L}{r^2}\right)}, &
t_{44}  = {\D \frac{p_r}{\Omega_r}}\end{array}\]
and
\[
h(r)= -\Phi'(r) + \frac{L^2}{r^3}, \qquad
p_r = \sqrt{2 (E - \Phi(r)) - \frac{L^2}{r^2}},
\]
where all functions are evaluated at ${\bf \bar{x}}$. 
Therefore the variance matrix in $({\bf x}, \,{\bf p})$ is
\begin{equation}
{\bf \sigma}_{\bf \omega} = ({\bf \Theta}(t) 
{\bf T}^{-1})^{\dagger} 
{\bf \sigma^0}_w({\bf \Theta}(t) {\bf T}^{-1}),
\end{equation}
so that, by substituting 
\[
{\bf \sigma_{\omega}} \!\!\!=\!\!\! \left[\begin{array}{cccc}
  \sigma_{11}  & 
  \sigma_{11}A & 
  \sigma_{11}B & 
	 \sigma_{11}C \\
   \{1,2\} & \sigma_{11}A^{2} + \sigma_{22}D^{2} +  
\sigma_{44} t_{42}^{2} & 
A \sigma_{11}B + D \sigma_{22}E +  
t_{42} \sigma_{44} t_{43} & 
 A \sigma_{11}C + D \sigma_{22}F +  t_{42} \sigma_{44} t_{44} \\
\{1,3\} &\{2,3\} & 
  \sigma_{11}B^{2} +  \sigma_{22}E^{2} +  \sigma_{33} +  
\sigma_{44} t_{43}^{2} & 
B \sigma_{11}C + E \sigma_{22}F +  t_{43} \sigma_{44} t_{44} \\ 
  \{1,4\} & \{2,4\} & \{3,4\} &
 \sigma_{11} C^{2} +  \sigma_{22}F^{2} +  \sigma_{44} t_{44}^{2} 
\end{array}\right] \]
and where
\begin{eqnarray*}
 A \!\!\!& = &\!\!\! t_{12} -\Omega'_{12} t_{42} t,  \qquad
 B =  t_{13} -\Omega'_{11}t - \Omega'_{12} t_{43}t, \qquad
 C =  t_{14}  -\Omega'_{12} t_{44}t,  \\
 D \!\!\!& = &\!\!\!  t_{22}  -\Omega'_{22} t_{42}t, \qquad
 E = t_{23}  -\Omega'_{12}t - \Omega'_{22} t_{43} t,\qquad
 F = t_{24}  -\Omega'_{22} t_{44}t.
\end{eqnarray*}
In general, one is more interested in the characteristics of the debris
in velocity space, rather than in momenta space. 
Thus we transform the variance matrix according to
$\sigma_{\varpi} = T_{p\rightarrow v}^{\dagger}
\sigma_{\omega} T_{p\rightarrow v}$, with
\begin{equation} 
{\bf T}_{p\rightarrow v} = \left[\begin{array}{cccc}1 & 0 & 0 & 0\\
0 & 1 & 0 & 0 \\ 0 & v_\psi & r & 0 \\
 0 & 0 & 0 & 1 \end{array}\right].
\end{equation}

The diagonalization of the variance matrix $\sigma_{\varpi}$
yields the values of the dispersions along the principal axes
and their orientation: two of its eigenvalues
increase with time, whereas the other two decrease with time. 
To understand the directly observable properties of the debris
we examine what happens around a particular point $\bar{\bf x}(t)$
in configuration space located on the mean orbit of the system. 
This is equivalent to
studying the velocity submatrix 
\[
{\bf \sigma_{\varpi}}(v) = \left[\begin{array}{cc}
  r^2 (\sigma_{11}B^{2} +  \sigma_{22}E^{2} +  \sigma_{33} +  
\sigma_{44} t_{43}^{2}) & 
r (B \sigma_{11}C + E \sigma_{22}F +  t_{43} \sigma_{44} t_{44}) \\
 \{1,2\} & 
 \sigma_{11} C^{2} +  \sigma_{22}F^{2} +  \sigma_{44} t_{44}^{2}
\end{array}\right].  
\]

For example, by diagonalising the matrix ${\bf \sigma_{\varpi}}(v)$ 
we obtain the directions of the 
principal axes of the velocity ellipsoid at the point $\bar{\bf x}(t)$,
and their dispersions.
Its eigenvalues are the roots of $\det[{\bf \sigma_{\varpi}}(v) -
\lambda {\bf {\cal I}}] = 0$. For $t \gg t_{\rm orb}$
\begin{eqnarray}
\lambda_i \!\!&=&\!\! \frac{t^2}{2}\left\{r^2 \left[\sigma_{11} 
\left(\Omega'_{11} -
\frac{\Omega'_{12}}{\Omega_r}\left(\Omega_\psi -
\frac{L}{r^2}\right)  \right)^2 + 
\sigma_{22} \left(\Omega'_{12} -
\frac{\Omega'_{22}}{\Omega_r}\left(\Omega_\psi -
\frac{L}{r^2}\right)  \right)^2 \right]+  \right. \nonumber \\
&& \,\, \left.\left(\frac{p_r}{\Omega_r}\right)^2 \left[\sigma_{11} 
{\Omega'_{12}}^2 + 
\sigma_{22} {\Omega'_{22}}^2\right] \pm \sqrt{R}\right\}, 
\end{eqnarray}
for $i=1,2$, and where
\begin{eqnarray}
R\!\!\!\!&=&\!\!\!\!\left\{r^2 \left[\sigma_{11} 
\left(\Omega'_{11} -
\frac{\Omega'_{12}}{\Omega_r}\left(\Omega_\psi -
\frac{L}{r^2}\right)  \right)^2 + 
\sigma_{22} \left(\Omega'_{12} -
\frac{\Omega'_{22}}{\Omega_r}\left(\Omega_\psi -
\frac{L}{r^2}\right)  \right)^2 \right] -  \right. \nonumber \\
&& \, \left. \left(\frac{p_r}{\Omega_r}\right)^2 \left[\sigma_{11} 
{\Omega'_{12}}^2 + 
\sigma_{22} {\Omega'_{22}}^2\right]\right\}^2 \nonumber \\
&+&\!\! 4 r^2 
\left(\frac{p_r}{\Omega_r}\right)^2 \left[\sigma_{11} \Omega'_{12}
\left(\Omega'_{11} -
\frac{\Omega'_{12}}{\Omega_r}\left(\Omega_\psi -
\frac{L}{r^2}\right)\right) + \sigma_{22} \Omega'_{22} \left(\Omega'_{12} -
\frac{\Omega'_{22}}{\Omega_r}\left(\Omega_\psi -
\frac{L}{r^2}\right)  \right)\right]^2.
\end{eqnarray}
Therefore 
\begin{eqnarray}
\lambda_1 \lambda_2 \!\!\!& = &\!\!\!
\left[t^2 (\Omega'_{11} \Omega'_{22}-{\Omega'_{12}}^2)
t_{44}\right]^2 \sigma_{11} \sigma_{22} r^2, \nonumber \\
\lambda_1 + \lambda_2 \!\!\!& = &\!\!\! t^2 r^2 
\left[ \sigma_{11} (\Omega'_{11} + 
\Omega'_{12} t_{43})^2 + \sigma_{22} (\Omega'_{12} + 
\Omega'_{22} t_{43})^2 \right] + t^2 \left[ \sigma_{11}{\Omega'_{12}}^2 +
 \sigma_{22}{\Omega'_{22}}^2\right] t_{44}^2. \nonumber
\end{eqnarray}
Since $\sigma(v_i) = \sqrt{1/\lambda_i}$ 
both  velocity dispersions decrease on the average as $1/t$.
The principal axes of the ellipsoid rotate as time 
passes by, not being coincident with any particular direction. 

\subsection{3-D treatment} 

As we discussed in Section~(4.2), the problem of the disruption of the system
and its evolution in phase-space is really a 3-D problem, since our initial
satellite had a finite width in all directions. Since we just discussed in
great detail what happens in the 2-D case and the way of proceeding once more
dimensions are added is the same, we will simply outline our main results,
focusing on what happens to the velocity submatrix.

If we assume that the system had initially a diagonal variance
matrix in action-angle variables, the velocity 
submatrix at time $t$ is
\begin{equation}
\label{eq:S_1}
{\bf \sigma_{\varpi}}(v) = {\bf T}^\dagger_{p\rightarrow v} {\bf S}(t) \,
{\bf T}_{p\rightarrow v},
\end{equation}
with 
\begin{equation}
\label{eq:S_2}
{\bf S}(t) = ({\bf W}_{ JJ} {\bf W}_{qJ}^{-1} - 
t\, {\bf \Omega'} {\bf W}_{ qJ}^{-1})^\dagger  \sigma_{\phi}^{0} 
({\bf W}_{JJ} {\bf W}_{ qJ}^{-1} - 
t \,{\bf \Omega'} {\bf W}_{ qJ}^{-1}) + 
{{\bf W}_{qJ}^{-1}}^\dagger \sigma_{J}^{0} {\bf W}_{qJ}^{-1},
\end{equation}
where ${\bf W}_{JJ}$ is the matrix whose elements are the second derivatives of
the characteristic function with respect to the actions,${\bf W}_{qJ}$ the
matrix that contains the second derivatives of $W$ with respect to the 
coordinates ${\bf q}$ and the actions ${\bf J}$, and 
$\Omega'_{ij} = \partial \Omega_i/
\partial J_j$. Note that, since the potential is spherical
${\bf \Omega'}$ and ${\bf W}_{JJ}$ have two equal rows. 
The initial variance matrix in action-angle space
\[
{\bf \sigma^0}_{w} = \left[\begin{array}{cc} {\bf \sigma}_{\phi}^0 & 
{\bf 0} \\
 {\bf 0} & {\bf \sigma}_{J}^0 \end{array} \right].
\]
We can  compute the density at a later time at the point $\bar{\bf x}(t)$
located on the mean orbit of the system
by integrating
\[
f({\bf {\bar x}}, {\bf v}, t) = f_0 \exp{\left[-\frac{1}{2}
{\bf \Delta}_\varpi^{\dagger}(t) {\bf \sigma}_{{\bf \varpi}}(t) 
{\bf \Delta}_\varpi(t)\right]},
\]
with respect to the velocities using 
the submatrix ${\bf \sigma_{\varpi}}(v)$ 
\[
\rho({\bf {\bar x}},t) = \int_{\Delta v_r}
\int_{\Delta v_\theta} \int_{\Delta v_\varphi} dv_\varphi\,dv_\theta \,dv_r \,
f({\bf \bar{x}}, {\bf v},t).
\]
In the principal axes frame
\begin{equation} 
\label{eq:rho_sigma}
\rho({\bf {\bar x}},t) = f_0 (2 \pi)^{3/2} 
\sigma_{v_1}(t) \sigma_{v_2}(t) \sigma_{v_3}(t) 
{\rm Erf}\!\!\left[\frac{a_{1}}{\sqrt{2} 
\sigma_{v_1}(t)}
\right]
{\rm Erf}\!\!\left[\frac{a_{2}}{\sqrt{2} \sigma_{v_2}(t)}\right]
{\rm Erf}\!\!\left[\frac{a_{3}}{\sqrt{2} \sigma_{v_3}(t)}\right],
\end{equation}
with $a_{1}, a_2, a_3$ the boundaries of the integration volume.
For $t \gg t_{\rm orb}$ the error function tends to $1$,
and therefore 
\begin{equation}
\rho(\bar{\bf x},t) =  (2 \pi)^{3/2} f_0 \,
\sigma_{v_1}(t) \sigma_{v_2}(t) \sigma_{v_3}(t),
\end{equation}
that is equivalent to
\begin{equation}
\label{eq:rho_lambdas}
 \rho(\bar{\bf x},t) =  (2 \pi)^{3/2} f_0/\sqrt{\lambda_1 \lambda_2 \lambda_3},
\end{equation}
where the $\lambda$'s are the eigenvalues of $\sigma_{\varpi}(v)$. With
simple algebra it can be shown that 
\begin{equation}
\label{eq:lambdas_det}
\lambda_1 \lambda_2 \lambda_3 = \det \sigma_{\varpi}(v),
\end{equation}
which is readily computable from Eqs.~(\ref{eq:S_1}) and ~(\ref{eq:S_2})
\begin{equation}
\label{eq:sigma_vf}
\det \sigma_{\varpi}(v) = (\det {\bf T}_{p\rightarrow v})^2 
(\det {\bf W}_{qJ}^{-1})^2  \det[({\bf W}_{JJ} - t\, {\bf \Omega'})^\dagger 
{\bf \sigma}_{\bf \phi}^{0} 
({\bf W}_{JJ} - t\, {\bf \Omega'}) +  {\bf \sigma}_{J}^{0}], 
\end{equation}
where 
\begin{equation}
\label{eq:t:p-v}
\det {\bf T}_{p\rightarrow v} = r^2 \sin\theta
\end{equation}
and 
\begin{equation}
\label{eq:det_t4}
\det {\bf W}_{qJ}^{-1} = \frac{p_r}{\Omega_r} \frac{ p_\theta}{L}.
\end{equation}
The remaining determinant in Eq.~(\ref{eq:sigma_vf}) for $t \gg t_{\rm orb}$ is
\[ \sigma_{33} (\sigma_{11} + \sigma_{22})(\sigma_{44} + \sigma_{55})
 (\Omega'_{11} \Omega'_{33} - {\Omega'_{13}}^2)^2 t^4,\] 
so that finally
\begin{equation}
\rho({\bf {\bar x}},t) = \frac{(2 \pi)^{3/2} f_0}
{|\Omega'_{11} \Omega'_{33}-{\Omega'_{13}}^2|}  \frac{\Omega_r L}
{\sqrt{\sigma_{33}(\sigma_{11}+\sigma_{22}) (\sigma_{44} + \sigma_{55})}}
 \frac{1}{r^2 \sin\theta |p_r p_\theta|}\frac{1}{t^2}.
\end{equation}
Let us recall that $\sigma_{ii} = 1/\sigma_{\phi_i}^2$ for $i=1..3$ and
$\sigma_{ii} = 1/\sigma_{J_j}^2$ for $i = j + 3 = 4..6$.

\section{Axisymmetric Eddington Potential}

To exemplify and understand how the rupture of the spherical symmetry 
affects the characteristic scales of the system, we take a very simple
Eddington  potential 
$\Phi(r,\theta) = \Phi_1(r) + \eta(\beta \cos{\theta})/r^2$
(Lynden-Bell 1962; 1994) which is separable in spherical coordinates.
The third integral for this class of potentials is
$I_3 = \frac{1}{2} L^2 + \eta(\beta \cos{\theta})$. The actions are
computed from:
\begin{eqnarray}
J_\varphi \!\!\!& = &\!\!\! L_z, \\
J_\theta \!\!\!& = &\!\!\! \frac{1}{2\pi}\oint d\theta \,
\sqrt{2 (I_3 - \eta(\theta)) - \frac{J_\varphi^2}
{\sin^2{\theta}}}, \\
J_r \!\!\!& = &\!\!\! \frac{1}{2\pi}\oint dr \, \sqrt{2 (E - \Phi_1(r)) - 
\frac{2I_3}{r^2}}.
\end{eqnarray}

The procedure outlined in Section 4.1 and Appendix A can also be applied to a
system moving in this type of potentials. In particular we are interested
in the behaviour of the density. By virtue of the previous discussion we
only need to find the determinant of the variance matrix as in 
Eq.~(\ref{eq:sigma_vf}), for this potential. Since Eqs.~(\ref{eq:t:p-v}) and
(\ref{eq:det_t4}) remain unchanged, we only focus on
$\det[({\bf W}_{JJ}{\bf W}_{qJ}^{-1} 
- t {\bf \Omega'}{\bf W}_{qJ}^{-1})^\dagger 
{\bf \sigma}_{\bf \phi}^{0} ({\bf W}_{JJ}{\bf W}_{qJ}^{-1} - 
t{\bf \Omega'}{\bf W}_{qJ}^{-1}) +  {\bf \sigma}_{J}^{0}]$. 
For $t \gg t_{\rm orb}$ the term with ${\bf \Omega'}$ will dominate 
with respect to ${\bf W}_{JJ}$, and the product 
$t^2 ({\bf \Omega'}{\bf W}_{qJ}^{-1})^\dagger {\bf \sigma}_{\bf \phi}^{0}
 {\bf \Omega'}{\bf W}_{qJ}^{-1}$ 
will dominate over  ${\bf \sigma^0}_{J}$\footnote{This does not
hold for the spherical case because $\det[(\Omega'{\bf W}_{qJ}^{-1})
^\dagger {\bf \sigma}_{\bf \phi}^{0} \Omega' {\bf W}_{qJ}^{-1}] \propto
\det \Omega' \equiv 0$}.  Therefore
\begin{equation}
\label{eq:ap_dets}
\det \sigma_{\varpi}(v) = 
(\det {\bf T}_{p\rightarrow v})^2 
(\det {\bf W}_{qJ}^{-1})^2 \nonumber (\det {\bf \Omega'}\,t)^2 
\det{\bf \sigma}_{\bf \phi}^{0},
\end{equation}
and so the density at the point ${\bf {\bar x}}$ at time $t$ is
\begin{equation}
\rho({\bf{\bar x}},t) =  \frac{(2 \pi)^{3/2} f_0}
{\sqrt{\det{\bf \sigma}_{\bf \phi}^{0}}}
\frac{1}{|\det{\bf \Omega'}|} 
\frac{\partial I_3}{\partial J_\theta} \frac{\Omega_r}
{r^2 \sin\theta |p_r p_\theta|}\frac{1}{t^3}.
\end{equation}

This expression is valid for a satellite described initially
by a Gaussian distribution. The variance matrix at $t=t^0$ may be
\begin{enumerate}
\item diagonal in action-angle variables: \[
\det{\bf \sigma}_{\bf \phi}^{0} = 1/(\sigma_{\phi_1} \sigma_{\phi_2}
\sigma_{\phi_3})^2, \]
\item diagonal in configuration-velocity space:
\begin{eqnarray}
{\det{\bf \sigma}_{\bf \phi}^{0}}&=& 
\frac{p^2_\theta p^2_r}
{\Omega_r^2 ({\partial I_3}/{\partial J_\theta})^2}
\frac{1}{\sigma_\varphi^2
\sigma_v^2}\left\{\frac{1}{\sigma_\theta^2}\left[
W_{rr}^2 + \frac{{v_\varphi}^2 + {v_\theta}^2}{r^2}\right] 
\right. \nonumber \\
&+& \left. \frac{1}{\sigma_v^2}
\left[W_{rr}^2 \frac{W_{\theta\theta}^2}{r^2} 
+ v_\varphi^2 \left( \frac{\cos^2 \theta}{\sin^2 \theta}W_{rr}^2 +
\left(\frac{v_\theta}{r} \frac{\cos \theta}{\sin \theta} -
\frac{W_{\theta\theta}}{r^2} \right)^2\right)\right]\right\}, \nonumber 
\end{eqnarray}
where all functions are evaluated at $({\bf \bar{x}^0}, {\bf \bar{v}^0})$, 
and
\[ W_{\theta\theta} = \frac{h_\theta}{p_\theta}, \qquad 
h_\theta = -\eta'(\theta) + 
J_\varphi^2 \frac{\cos\theta}{\sin^3 \theta},\]
and 
\[ W_{rr} = \frac{h_r}{p_r}, \qquad 
h_r = -\Phi'_1(r) + \frac{2 I_3}{r^3}.\]
The expression for the determinant of the angle submatrix at $t=0$
may be simplified
if the satellite is initially close to a turning point of the orbit.
In this case the term $W_{\theta\theta}^2 W_{rr}^2$ will be dominant
and 
\[{\det{\bf \sigma}_{\bf \phi}^{0}} = \left[\frac{h_\theta(\theta^0) h_r(r^0)}
{\Omega_r \, r^0} \left(\frac{\partial I_3}{\partial J_\theta}\right)^{-1}
\frac{1}{\sigma_\varphi \sigma_v^2}\right]^2.\]
\end{enumerate}

Note that the main differences with the spherical case are
\begin{itemize}
\item the time dependence: $t^3$ instead of $t^2$ because of the 
increase in the dimensionality of the problem; 
\item the dependence on the derivatives of the basic frequencies of motion:
 the same functional dependence $\det{\bf \Omega'}$, but now with three
independent frequencies and derivatives;
\item the inclusion of the term $\partial I_3/\partial J_\theta$, which for
the spherical case is simply $L$;
\item the form of $p_\theta = \sqrt{ 2 (I_3 - \eta(\theta)) - 
J_\varphi^2/\sin^2\theta}$, which also includes the angular dependence  of the 
potential. 
\end{itemize}

\section{Axisymmetric St\"ackel Potential}

In this section we collect some basic properties of St\"ackel potentials
and derive the density behaviour as a function of time, as in previous 
sections, from Liouville's Theorem and the evolution 
of the system in action-angle variables. Further details on St\"ackel
potentials can be found in de Zeeuw (1985).

Let us first introduce spheroidal coordinates $(\lambda, \nu, \varphi)$,
where $\varphi$ is the azimuthal angle in the usual cylindrical coordinates
$(R, z, \varphi)$, and $\lambda$ and $\nu$ are the two roots for $\tau$ of
\begin{equation}
\frac{R^2}{\tau - a^2} + \frac{z^2}{\tau - c^2} = 1,
\end{equation}
where $c^2 \le \nu \le a^2 \le \lambda$. A potential is of St\"ackel form
if it can be expressed as
\begin{equation}
\label{eq:pot_staeckel}
V = - \frac{(\lambda - c^2) G(\lambda) - (\nu - c^2) G(\nu)}{\lambda - \nu},
\end{equation}
where $G(\tau)$ is an arbitrary function ($\tau = \lambda, \nu$). In this case,
the Hamiltonian becomes
\begin{equation}
H = \frac{p^2_\lambda}{2 P^2} + \frac{p^2_\nu}{2 Q^2}+ \frac{p^2_\varphi}
{2 R^2} +
V(\lambda, \nu)
\end{equation}
where the functions $P$ and $Q$ are 
\begin{equation}
P^2 = \frac{\lambda - \nu}{4 (\lambda - a^2) (\lambda - c^2)}, 
\qquad Q^2 = \frac{\nu - \lambda}{4 (\nu - a^2) (\nu - c^2)}.
\end{equation}
Three isolating integrals of motion 
can be found ($E$, $I_2$, $I_3$), and the system is separable
since the equations of motion can be written as
\begin{equation}
p^2_{\tau} = \frac{1}{2 (\tau - a^2)} \left[G(\tau) - \frac{I_2}{\tau - a^2} -
\frac{I_3}{\tau - c^2}\right], \,\, \,\, \tau = \lambda, \nu,
\end{equation}
and 
\begin{equation}
p_\varphi = L_z = \sqrt{2 I_2}.
\end{equation}

To represent the Galaxy we may choose
a superposition of two St\"ackel potentials: a disk plus a halo component
\begin{equation}
V = k V_{\rm disk} + (1 - k) V_{\rm halo}, 
\end{equation}
where $k$ represents the mass fraction of the
disk with respect to the total mass of the Galaxy. Since the coordinates
used for the halo and the disk have to be the same, this introduces a relation
between the characteristic parameters $(a_{\rm d}, c_{\rm d})$ and 
$(a_{\rm h}, c_{\rm h})$ of
the St\"ackel potentials. It can be shown that 
the potential 
\begin{equation}
V(\lambda, \nu, q) = - GM \left[\frac{k}{\sqrt{\lambda} + \sqrt{\nu}} +
\frac{1- k}{\sqrt{\lambda+q} + \sqrt{\nu+q}}\right]
\end{equation}
where $q$ is related to the flattening of the halo component, 
provides a good description yielding a flat rotation curve
with similar properties to that of our Galaxy (Batsleer \& Dejonghe 1994).
The function $G(\tau)$ in Eq.~(\ref{eq:pot_staeckel}) is
\begin{equation}
G(\tau) = G M \left[ \frac{k}{\sqrt{\tau} + c} + 
\frac{1-k}{\sqrt{\tau + q} + c} \right]
\end{equation}
For the characteristic parameters we choose 
$a_{\rm d} = 2$, $c_{\rm d} = 1$, $a_{\rm h}/c_{\rm h} =  1.01$
(giving a rather spherical halo), $k = 0.12$ and $M = 5 \times 10^{11} \sm$.  

In order to obtain the evolution of the mean density of debris
as a function of time in a St\"ackel potential we use the results of
Section 4 and of Appendix A and B. From Eqs.~(\ref{eq:rho_lambdas}) 
and (\ref{eq:lambdas_det}) the density is
proportional to the determinant of the velocity submatrix. Since 
the Hamiltonian is separable in spheroidal coordinates, to obtain
the density in cylindrical (or spherical) coordinates we need to multiply
Eq.~(\ref{eq:ap_dets}) by the determinant of the matrix that 
performs the transformation between the two sets of coordinates. Thus
\begin{equation}
\det{\bf \sigma}_\varpi(v)  = \left[\det {\bf T}_{p\rightarrow v} 
\det {\bf T}_{p_\tau 
\rightarrow p_{cyl}} \det {\bf W}_{qJ}^{-1}
\det{\bf \Omega'}t\right]^2 \det {\bf \sigma}_{\bf \phi}^{0},
\end{equation}
where
\begin{equation}
\det {\bf T}_{p\rightarrow v} = R, \qquad
\det {\bf T}_{p_\tau \rightarrow p_{cyl}}  \det {\bf W}_{qJ}^{-1} =
\D\frac{(\nu - \lambda) v_\lambda v_\nu}{\D\Omega_\nu \D\frac{\D\partial I_3}
{\D\partial J_\lambda} - \Omega_\lambda \frac{\partial I_3}{\partial J_\nu}}. 
\end{equation}
The mean density at time $t$ at the point
$\bar{\bf x}$ on the mean orbit of the system becomes
\begin{equation}
\label{eq:rho_staeckel}
\rho({\bf{\bar x}},t ) = \frac{(2 \pi)^{3/2} f_0}
{\sqrt{\det{\bf \sigma_\phi^0}}}
\frac{1}{|v_\lambda v_\nu| |\lambda-\nu| R}
\frac{\D\left|\Omega_\nu \frac{\partial I_3}{\partial J_\lambda} - 
\Omega_\lambda \frac{\partial I_3}{\partial J_\nu}\right|}{|\det{\bf \Omega'}|}
\frac{1}{t^3}.
\end{equation}
This expression is valid for a satellite described initially
by a Gaussian distribution. The variance matrix at $t=t^0$ may be
\begin{enumerate}
\item diagonal in action-angle variables: \[
\det{\bf \sigma}_{\bf \phi}^{0} = 1/(\sigma_{\phi_1} \sigma_{\phi_2}
\sigma_{\phi_3})^2, \]
\item diagonal in configuration-velocity space. If
the satellite is initially close to a turning point of the orbit
then
\begin{equation}
\label{eq:det_sigma_st}
{\det{\bf \sigma}_{\bf \phi}^{0}} = \left[\frac{h_\lambda 
h_\nu}{\D\left|\Omega_\nu \frac{\partial I_3}{\partial J_\lambda} - 
\Omega_\lambda \frac{\partial I_3}{\partial J_\nu}\right|} 
\frac{\lambda-\nu}{P^3 Q^3}
\frac{1}{\sigma_\varphi \sigma_v^2}\right]^2,
\end{equation}
where all functions are evaluated at $({\bf \bar{x}^0}, {\bf \bar{v}^0})$, 
and
\[ h_\tau = 2 p_\tau \frac{\partial p_\tau}{\partial \tau}, 
\qquad \tau = \lambda, \nu.\]
\end{enumerate}

\label{lastpage}

\end{document}